\documentclass{aa}
\usepackage[varg]{txfonts}
\usepackage{graphicx}
\usepackage{natbib}
\usepackage{stfloats}
\usepackage[colorlinks=true, linkcolor=blue, citecolor=blue, filecolor=blue, urlcolor=blue]{hyperref}
\bibpunct{(}{)}{;}{a}{}{,}
\usepackage{orcidlink}
\usepackage{array}
\usepackage{enumitem}
\usepackage{placeins}
\DeclareSymbolFont{upgreek}{U}{eur}{m}{n}
\DeclareMathSymbol{\umu}{0}{upgreek}{"16}
\newcolumntype{x}[1]{>{\centering\arraybackslash\hspace{0pt}}p{#1}}

\title{Circular polarization as a probe of cloud properties and asymmetries in giant exoplanet atmospheres}
\titlerunning{Circular polarization as a probe of cloud properties and asymmetries}

\author{M. B. Michaelis\orcidlink{0009-0007-6337-7023} \and S. Wolf\orcidlink{0000-0001-7841-3452}}
\authorrunning{Michaelis \& Wolf}
\institute{Institute of Theoretical Physics and Astrophysics, Kiel University, Leibnizstr. 15, 24118 Kiel, Germany\\\email{mmichaelis@astrophysik.uni-kiel.de}}
\date{Received <date> / Accepted <date>}
    
\abstract
    {Polarimetry is a promising method for characterizing clouds in exoplanetary atmospheres. For planets in the Solar System, circular polarization measurements complement linear polarimetry by providing additional information on cloud particle properties. As the disk-integrated circular polarization is zero for symmetric planets, observing intrinsic circular polarization of spatially unresolved exoplanets requires stable spatial asymmetries such as circumplanetary rings.
    }
    {We investigated the potential of circular polarization measurements at optical and near-infrared wavelengths to determine optical properties of cloud particles in the atmospheres of giant exoplanets and characterize asymmetries.
    }
    {For 20 selected cloud condensates that cover a large range of refractive indices, we calculated optical properties assuming Mie scattering theory. The circular polarization of starlight scattered by cloudy exoplanets was calculated with Monte Carlo radiative transfer simulations. To explain the connection between optical properties and planetary circular polarization, we derived an interpretative model of the first two scattering orders.
    }
    {Planetary hemispheres with atmospheres including cloud particles with a large imaginary part, $k$, of the refractive index show distinct circular polarization phase curves dominated by scattering first by gaseous molecules and then by cloud particles. The intrinsic degree of circular polarization, $P_\mathrm{c}$, is at most $3\cdot 10^{-4}$. When the cloud particles have a low $k$, they instead induce even smaller but more predictable circular polarization dominated by scattering solely by cloud particles. The predictability allows for a characterization of the underlying asymmetry. For example, this could be used to determine which half of the planet is covered by a circumplanetary ring.
    }
    {Circular polarization of starlight reflected by giant exoplanets is sensitive to cloud particle composition and large-scale asymmetries but remains a subtle signal. While promising for characterizing clouds under favorable conditions, practical detection requires technological advances in polarimetry and careful disentanglement from stellar background signals.
    }
    

\begin{document}

\maketitle
\nolinenumbers
\section{Introduction}
\label{sec:introduction}

    Eye-catching features of planetary atmospheres in the Solar System are
    clouds, such as terrestrial water clouds, sulfuric acid clouds on Venus \citep{young1973,hansen1974}, and different cloud layers in the Jovian atmosphere containing ammonia and ammonium hydrosulfide among various other condensates hidden in deeper layers \citep{atreya1999}.
    Similarly, clouds are predicted to form in the atmospheres of exoplanets \citep[e.g.,][]{sanchez_lavega2004,helling2019,roman2021}.
    Their presence blocks our view of the gaseous layers beneath, limiting our ability to characterize the atmosphere using transit or eclipse spectroscopy \citep[e.g.,][]{seager2000}. Indeed, absorption features appear muted or are obscured entirely in optical or near-infrared transmission spectra of many observed exoplanets \citep[e.g.,][]{charbonneau2002,tinetti2007,pont2008,pont2013,bean2010,berta2012,gibson2013,knutson2014,kreidberg2014,kreidberg2018,sing2016,espinoza2019}.
    
    Clouds also affect the phase angle dependence of the reflected light flux from exoplanets \citep[e.g.,][]{seager2000,heng2013,webber2015,dyudina2016,oreshenko2016}.
    Therefore, phase curve measurements at visible wavelengths 
    are of great value for determining the properties of the planet and its clouds \citep{garcia_munoz2015b,feng2018,garcia_munoz2018,vaughan2023,akinsanmi2024}.
    In Solar System research, polarimetric phase curves further constrain cloud particle properties.
    For instance, \citet{hansen1974} determined the composition, particle size distribution, and vertical structure of the Venusian clouds using linear polarization measurements performed by \citet{lyot1929}, \citet{coffeen1969}, \citet{dollfus1970}, and \citet{veverka1971}.
    \citet{kawata1978} interpreted circular polarization measurements conducted by \citet{kemp1971b}, \citet{swedlund1972}, and \citet{michalsky1974} and narrowed down the particle sizes and refractive indices of Jovian and Saturnian clouds compared to an analysis using only linear polarimetry.
    
    Consequently, polarimetry is a promising method for characterizing atmospheres of exoplanets \citep[e.g.,][]{rossi2022}.
    Numerical studies demonstrate that linear polarization measurements of starlight reflected by exoplanets constrain their 
    cloud composition and cloud particle size \citep[e.g.,][]{karalidi2011,bailey2018,lietzow2022}, or the fraction of the planet covered by clouds \citep{rossi2017,winning2024}.
    Recently, \citet{wiktorowicz2025} reported the discovery of linearly polarized flux of the hot Jupiter HD 189733b in the B band with an amplitude of $40\ \mathrm{ppm}$ of the stellar flux. Their measurements are in accordance with previous spectrometric detections of small $\mathrm{SiO_2}$ particles \citep{inglis2024}.
    In addition, \citet{wiktorowicz2025} detect circularly polarized flux 
    that is attributed to star-planet interactions, as the amplitude of $67\ \mathrm{ppm}$ exceeds that expected for a planetary atmosphere. For reference, the intrinsic circular polarization of sunlight reflected by the planets in the Solar System is three orders of magnitude below its linear polarization, even if the planetary hemispheres above and below the planetary scattering plane, which contains the observer and the centers of star and planet, are observed separately \citep[e.g.,][]{kemp1971b,swedlund1972,michalsky1974}.
    
    If a planet is mirror-symmetric with respect to the scattering plane, the circularly polarized fluxes of the two hemispheres have the same amplitude but opposite handedness, which is called the polar effect. As a consequence, the circular polarization of starlight reflected by unresolved planets or exoplanets is zero, unless an asymmetry between the hemispheres exists \citep[e.g.,][]{hansen1971c,hansen1974b,wolstencroft1976}. For instance, the Saturnian ring system partially occults one hemisphere of Saturn for an observer on Earth, eliminating the polar effect and resulting in a net circular polarization for unresolved measurements \citep{smith1983}.
    
    For ocean planets and Earth analogs, the circularly polarized flux is four magnitudes smaller than the linearly polarized flux \citep{garcia_munoz2015,rossi2018,groot2020}.
    Although polarimetry is currently more promising for giant exoplanets \citep{wiktorowicz2025}, we are not aware of any detailed numerical studies that discuss their circular polarization.
    
    This study investigates the potential to characterize cloud layers in hydrogen-dominated planetary atmospheres by observing the circular polarization that is due to the presence of a large-scale asymmetry 
    at wavelengths from $0.3\ \mathrm{\umu m}$ to $1\ \mathrm{\umu m}$, with a focus on $0.5\ \mathrm{\umu m}$. 
    Sect. \ref{sec:methods} introduces our framework for describing the polarization state and our calculations of the optical properties of the cloud compositions considered. 
    To facilitate the explanation of the connection between the optical properties of cloud particles and the circular polarization of light reflected by one hemisphere of an exoplanet, Sect. \ref{sec:semi_analytical_calculations} introduces an interpretative second-scattering-order model. 
    The findings are 
    compared to a model including all scattering orders and an atmosphere consisting of hydrostatic gaseous layers and a cloud layer in Sect. \ref{sec:results}.
    More realistic scenarios in which a nonzero circular polarization is expected, such as circumplanetary rings, and the observability of planetary circular polarization signals, are discussed in Sect.~\ref{sec:discussion}. We summarize our findings in Sect. \ref{sec:conclusion}.

\section{Methods}
\label{sec:methods}
    
    The polarization state of electromagnetic radiation is described with its Stokes vector, $\mathbf{S} = (I, Q, U, V)^T$, including the total intensity, $I$, linearly polarized intensities, $Q$ and $U$, and circularly polarized intensity, $V$ \citep[e.g.,][]{stokes1852,bohren1983}.
    The degree of linear polarization, $P_\mathrm{l}$, direction of linear polarization, $\chi$, and degree of circular polarization, $P_\mathrm{c}$, were calculated according to \citet{bohren1983}:
    \begin{equation}
        P_\mathrm{l} = \frac{\sqrt{Q^2 + U^2}}{I}, \quad
        \tan 2\chi = \frac{U}{Q}, \quad P_\mathrm{c} = \frac{V}{I}.
    \end{equation}
    Positive values of $V$ or $P_\mathrm{c}$ indicate that the electric field vector rotates counterclockwise for an observer facing the planet. Negative values indicate clockwise rotation \citep{kawata1978}.
    In Solar System research, it is common to use the sign-dependent degree of linear polarization, $P_\mathrm{s} = -P_\mathrm{l}\cos2(\chi-\psi)$, with the position angle of the scattering plane, $\psi$ \citep{zubko2019}.

    If the reference plane of the Stokes vector is rotated around the axis defined by the wave vector by an angle of $\Phi$, the Stokes vector, $\vec{S}_\mathrm{old}$, in the old reference system is transformed to the new system according to $\vec{S}_\mathrm{new} = \mathbf{L}(\Phi)\cdot \vec{S}_\mathrm{old}$, where the rotation matrix is defined by \citep[e.g.,][]{bohren1983}
    \begin{equation}
        \mathbf{L}(\Phi) = 
        \begin{pmatrix}
            1 & & & \\
            & \cos 2\Phi & \sin 2\Phi & \\
            & -\sin 2\Phi & \cos 2\Phi & \\
            & & & 1
        \end{pmatrix}.
        \label{eq:rotation_matrix}
    \end{equation}
    When radiation with Stokes vector $\vec{S}$ is scattered, the Stokes vector, $\vec{S}^\prime$, of the fraction of radiation that was scattered by a polar angle, $\Theta$, and an azimuthal scattering angle, $\Phi$, is \citep[e.g.,][]{van_de_hulst1948,bohren1983}
    \begin{equation}
        \vec{S}^\prime = \frac{\varpi}{4\pi}\mathbf{F}(\Theta,\Phi)\cdot\vec{S}.
        \label{eq:scattering_mueller_formalism}
    \end{equation}
    Here, the scattering plane is the reference plane of $\vec{S}$ and $\vec{S}^\prime$.
    The quantity $\varpi = C_\mathrm{sca}/C_\mathrm{ext}$ is the single scattering albedo of the particle ensemble that is calculated from the scattering and extinction cross sections, $C_\mathrm{sca}$ and $C_\mathrm{ext}$, respectively. $\mathbf{F}$ is the $4\times 4$ scattering matrix with matrix elements $F_{ij}$ \citep[e.g.,][]{bohren1983}.
    $\mathbf{F}$ is normalized such that the integral of $F_{11}$ with respect to the solid angle is $4\pi$ \citep{hansen1974b}.

    \begin{table*}
        \centering
        \caption{Considered cloud particle species.}
        \begin{tabular}{llllll}
            \hline\hline\\[-0.9em]
            Formula & Name & Optical properties reference & $n$ & $k$ & $\varpi$ \\ 
            \hline\\ [-0.9em]
            $\mathrm{Al_2O_3}$ & corundum & \citet{koike1995} & $1.58$ & $3.82\cdot 10^{-2}$ & $0.59$ \\ 
            $\mathrm{C}$ & graphite & \citet{draine2003} & $2.55$ & $1.20$ & $0.60$ \\ 
            $\mathrm{CaTiO_3}$ & perovskite & \citet{ueda1998} & $2.36$ & $2.74 \cdot 10^{-3}$ & $0.93$ \\ 
            $\mathrm{Cr}$ & chromium & Lynch \& Hunter in \citet{palik1991} & $2.61$ & $4.45$ & $0.79$ \\
            $\mathrm{Fe}$ & iron &  Lynch \& Hunter in \citet{palik1991} & $2.00$ & $3.07$ & $0.73$ \\ 
            $\mathrm{FeO}$ & wüstite & \citet{henning1995}\ \tablefootmark{(a)} & $2.34$ & $7.42\cdot 10^{-1}$ & $0.57$ \\ 
            $\mathrm{FeS}$ & troilite & \citet{pollack1994} & $1.36$ & $1.37$ & $0.61$ \\
            $\mathrm{Fe_2O_3}$ & hematite & A.H.M.J. Triaud, unpublished\ \tablefootmark{(a)} & $2.99$ & $7.80\cdot 10^{-1}$ & $0.60$ \\ 
            $\mathrm{Fe_2SiO_4}$ & fayalite & \citet{fabian2001}\ \tablefootmark{(a)} & $1.85$ & $1.22 \cdot 10^{-3}$ & $0.96$ \\ 
            $\mathrm{H_2O}$ & water ice & \citet{warren2008} & $1.31$ & $5.89 \cdot 10^{-10}$ & $1.00$ \\ 
            $\mathrm{KCl}$ & sylvite & Palik in \citet{palik1985} & $1.50$ & $7.60 \cdot 10^{-11}$ & $1.00$ \\
            $\mathrm{MgAl_2O_4}$ & spinel & Tropf \& Thomas in \citet{palik1991} & $1.72$ & $2.20 \cdot 10^{-13}$ & $1.00$ \\ 
            $\mathrm{MgFeSiO_4}$ & olivine & \citet{dorschner1995}\ \tablefootmark{(a)} & $1.77$ & $1.00\cdot 10^{-1}$ & $0.52$ \\
            $\mathrm{Mg_2SiO_4}$ & forsterite & \citet{jaeger2003}\ \tablefootmark{(a)} & $1.62$ & $1.70\cdot 10^{-4}$ & $0.99$ \\ 
            $\mathrm{MnS}$ & manganese sulfide & \citet{huffman1967} & $2.95$ & $2.72\cdot 10^{-3}$ & $0.93$ \\ 
            $\mathrm{Na_2S}$ & sodium sulfide & \citet{khachai2009} & $1.85$ & $1.64 \cdot 10^{-2}$ & $0.73$ \\ 
            $\mathrm{NH_3}$ & ammonia ice & \citet{martonchik1984} & $1.44$ & $2.64 \cdot 10^{-5}$ & $1.00$ \\ 
            $\mathrm{SiO}$ & silicon-monoxide & Philipp in \citet{palik1985} & $2.03$ & $4.37 \cdot 10^{-2}$ & $0.60$ \\ 
            $\mathrm{SiO_2}$ & quartz, silica & Philipp in \citet{palik1985} & $1.55$ & $5.71\cdot 10^{-6}$ & $1.00$ \\ 
            $\mathrm{ZnS}$ & zinc sulfide & Palik \& Addamiano in \citet{palik1985} & $2.42$ & $4.78 \cdot 10^{-6}$ & $1.00$ \\ 
            \hline
        \end{tabular}
        \tablefoot{
            $n$ and $k$ are the real and imaginary parts of the refractive index at a wavelength of $0.5\ \mathrm{\umu m}$. The single scattering albedo, $\varpi$,
            was calculated using the \emph{miex} algorithm \citep{wolf2004}.
            This is a subset of the optical dataset used by \citet{lietzow2022}. Most of their data was provided by \citet{kitzmann2018} as part of the \href{https://github.com/exoclime}{Exoclime} simulation package.
            \tablefoottext{a}{part of the \href{https://www2.astro.uni-jena.de/Laboratory/OCDB/}{Database of Optical Constants for Cosmic Dust, Laboratory Astrophysics Group of the AIU Jena}}
        }
        \label{tab:particle_species}
    \end{table*}
    
    Our planetary models include atmospheres composed of $\mathrm{H}_2$ molecules and cloud particles \citep[similar to][]{lietzow2022}. Hemispheric asymmetries are not included in our models. Instead, the planetary Stokes vector is integrated only on the planetary hemisphere above the planetary scattering plane. While this is not a realistic scenario, it still provides an approximate upper limit on the amplitude of a planetary circular-polarization signal. For instance, a circumplanetary ring could cover a large part but not the entirety of a hemisphere. A brief discussion of the influence of possible asymmetries is given in Sect. \ref{sec:discussion}.
    
    The Rayleigh scattering cross section of $\mathrm{H}_2$ was calculated according to \citet{sneep2005} based on the wavelength-dependent refractive index of $\mathrm{H}_2$ \citep{keady2000}. The Rayleigh scattering matrix is found in \citet{hansen1974b}. We assumed a depolarization factor of $\mathrm{H}_2$ of $0.02$ at all considered wavelengths \citep{penndorf1957}.
    Similar to \citet{lietzow2022}, absorption by gaseous molecules was ignored.
    
    To facilitate comparisons of planetary circular polarization phase curves presented in Sect. \ref{sec:semi_analytical_calculations} and \ref{sec:results} with linear polarization results by \citet{lietzow2022}, cloud particles were assumed to be spherical, homogeneous, and in the solid phase.
    Cross sections and scattering matrices of cloud particles were calculated according to Mie theory \citep{mie1908}
    using the \emph{miex} algorithm \citep{wolf2004} and averaged over the distribution of their radii $r$ \citep[see, e.g.,][]{martin1978,wolf2003,steinacker2013}, which was assumed to be a gamma distribution \citep{hansen1971c,hansen1974b}, 
    \begin{equation}
        N(r) \propto r^{(1 - 3v_\mathrm{eff})/v_\mathrm{eff}} e^{-r/(r_\mathrm{eff}v_\mathrm{eff})}
        \label{eq:size_distribution}
    ,\end{equation}
    with an effective radius of $r_\mathrm{eff} = 1\ \mathrm{\umu m}$ and an effective variance of $v_\mathrm{eff} = 0.1$ \citep[similar to][]{lietzow2022}. The wavelength range of interest from $0.3$ to $1\ \mathrm{\umu m}$ corresponds to size parameters $x_\mathrm{eff} = 2\pi r_\mathrm{eff}/\lambda$ \citep[e.g.,][]{van_de_hulst1957} between $21$ and $6$.
    For Mie and Rayleigh scattering, the scattering matrix is of the form \citep[e.g.,][]{van_de_hulst1957,bohren1983}
    \begin{equation}
        \mathbf{F}(\Theta) = \begin{pmatrix}
            F_{11} & F_{12} & & \\
            F_{12} & F_{22} & & \\
            & & F_{33} & F_{34} \\
            & & -F_{34} & F_{44}
        \end{pmatrix}.
        \label{eq:scattering_matrix}
    \end{equation}
    $\mathbf{F}$ does not depend on the azimuthal angle, $\Phi$.
    For Mie scattering by spherical particles, $F_{11}$ equals $F_{22}$, and similarly $F_{33}$ equals $F_{44}$ \citep[e.g.,][]{van_de_hulst1957,bohren1983}.

    Assuming spherical cloud particles is a common approximation for modeling exoplanetary atmospheres \citep[e.g.,][]{rossi2018,benneke2019,lietzow2022,mullens2024,wiktorowicz2025,batalha2026}.
    However, the polarization of the reflected light is sensitive to the shape of the scattering particles \citep{mishchenko1994}. For example, cuboid $\mathrm{KCl}$ particles induce polarization directions perpendicular to those expected for spherical particles \citep{hamill2024}.
    The presence of irregularly shaped aggregate particles in the atmospheres of Jupiter and Titan is inferred from polarimetric observations \citep{west1991,dlugach2005}. Consequently, nonspherical cloud particles are also expected to form in exoplanetary atmospheres \citep[e.g.,][]{samra2020,chubb2024,hamill2025}. Modeling nonspherical particles is beyond the scope of this study. However, possible effects of irregularly shaped particles are discussed qualitatively in Sect. \ref{sec:discussion}.

    We used the set of the cloud particle species considered by \citet{lietzow2022}, but excluded $\mathrm{CH_4}$, $\mathrm{MgO}$, $\mathrm{MgSiO_3}$, $\mathrm{NaCl}$, and $\mathrm{TiO_2}$, as their refractive indices are very similar to $\mathrm{H_2O}$, $\mathrm{MgAl_2O_4}$, $\mathrm{SiO_2}$, or $\mathrm{ZnS}$, respectively (see Table \ref{tab:particle_species}).
    The computed scattering matrix elements are visualized in App. \ref{app:optical_properties}.
    Depending on the temperature profile of a giant exoplanet, different materials are expected to form clouds in the upper layers of the atmosphere \citep{sanchez_lavega2004,allard2012,morley2013}.
    Ammonia and water ice are expected in atmospheres of cool gas giants at temperatures lower than $350 \ \mathrm{K}$ \citep[e.g.,][]{atreya1999,sanchez_lavega2004}. Graphite clouds are expected in atmospheres of rocky or gaseous exoplanets with high atmospheric C/O ratios at gas temperatures of around $500\ \mathrm{K}$ \citep{herbort2022,li2025}.
    Other species discussed in this study are expected to condense in the atmospheres of hot exoplanets \citep{wakeford2015,kitzmann2018,helling2019,roman2021}, with condensation temperatures reaching from about $700\ \mathrm{K}$ for $\mathrm{KCl}$ to more than $1800\ \mathrm{K}$ for corundum ($\mathrm{Al_2O_3}$) \citep[e.g.,][]{lodders2002,wakeford2017,batalha2026}.

\section{Second scattering order calculations}
\label{sec:semi_analytical_calculations}

    The matrix element $F_{14}$ equals zero for Mie and Rayleigh scattering (see Eq. \ref{eq:scattering_matrix}). Thus, calculations that include two scattering events are needed to derive an interpretative model describing the connection between features of the scattering matrix and planetary circular polarization \citep[e.g.,][]{wolstencroft1976,kawata1978}.

\begin{figure}
        \centering
        \includegraphics{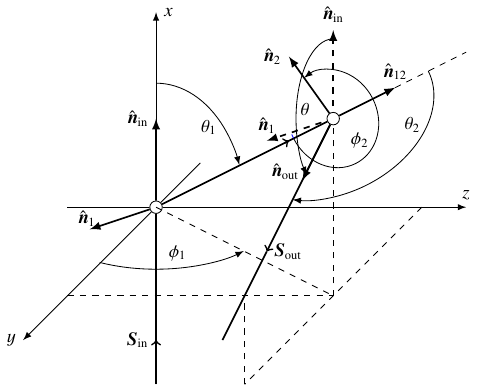}
        \caption{Double scattering geometry. Two particles are shown as circles. $\hat{\vec{n}}_\mathrm{in}$ (incoming radiation, $\vec{S}_\mathrm{in}$), $\hat{\vec{n}}_{12}$ (after first scattering), and $\hat{\vec{n}}_\mathrm{out}$ (outgoing radiation, $\vec{S}_\mathrm{out}$) define the radiation path. $\hat{\vec{n}}_1$ and $\hat{\vec{n}}_2$ are the normals of the first and second scattering plane, respectively. $\theta_1, \phi_1,$ and $\theta_2, \phi_2$ are the polar and azimuthal scattering angles of the first and second scattering event, respectively. The observer is located in the $xy$ plane at a positive $y$ coordinate. Hence, the overall azimuthal scattering angle is $\phi = 0$. $\theta$ is the overall polar scattering angle.}
        \label{fig:two_particle_model}
    \end{figure}

    For double scattering, the incoming radiation is scattered into a random direction once before it is scattered toward the observer in a second scattering event. We assume that both scatterings are described by a scattering matrix, $\mathbf{F}(\Theta)$, of the form in Eq. \eqref{eq:scattering_matrix} and a single scattering albedo, $\varpi$. A sketch of the geometry is shown in Fig. \ref{fig:two_particle_model}.
    Here, $\theta_1$ and $\phi_1$ are the polar and azimuthal scattering angles of the first scattering event, respectively. The observer is located at a position defined by an overall scattering angle, $\theta$.
    The polar and azimuthal scattering angles of the second scattering event, $\theta_2$ and $\phi_2$, respectively, are functions of $\theta$, $\theta_1$, and $\phi_1$ (see App. \ref{app:second_scattering_angles}). The angle between the second scattering plane and the reference plane of the observer is $\beta$.
    Because the disk-integrated linear and circular polarization of the quiet Sun are below $1\ \mathrm{ppm}$ \citep{kemp1987}, the incident stellar radiation is assumed to have a Stokes vector of $\vec{S}_\mathrm{in} = (I_0, 0,0,0)^T$.
    
    According to Eq. \eqref{eq:scattering_mueller_formalism} and using the rotation matrix given in Eq. \eqref{eq:rotation_matrix}, the Stokes vector of radiation that was first scattered in a direction given by the angles $\theta_1, \phi_1$ before being scattered toward an observer in the direction given by $\theta$ is
    \begin{equation}
        \vec{S}_\mathrm{out}(\theta, \theta_1, \phi_1) = \left(\frac{\varpi}{4\pi}\right)^2\,\mathbf{L}(\beta)\cdot\mathbf{F}(\theta_2)\cdot\mathbf{L}(\phi_2)\cdot\mathbf{F}(\theta_1)\cdot\mathbf{L}(\phi_1)\cdot\vec{S}_\mathrm{in}.
        \label{eq:S_out_main}
    \end{equation}
    Multiplications with $\mathbf{L}(\phi_1)$ and $\mathbf{L}(\beta)$ do not change the Stokes parameters $I_\mathrm{out}$ and $V_\mathrm{out}$. 
    \citet{kawata1978} derived an equivalent equation for the case of $\varpi = 1$. 
    
    $\vec{S}_\mathrm{out}$ must be integrated on all first scattering directions. However, the integral of the circularly polarized component
    \begin{equation}
        \int_0^{2\pi}\int_0^\pi V_\mathrm{out}(\theta,\theta_1,\phi_1)\sin\theta_1\,\mathrm{d}\theta_1\,\mathrm{d}\phi_1 = 0
        \label{eq:anti-symmetry}
    \end{equation}
    is always zero due to the anti-symmetry (see App. \ref{app:local_azimuthal_effect})
    \begin{equation}
        V_\mathrm{out}(\theta,\theta_1,\phi_1) = -V_\mathrm{out}(\theta,\theta_1,2\pi - \phi_1).
        \label{eq:anti-symmetry-V}
    \end{equation}
    To get an informative result, including the geometry of the problem in the calculations is therefore unavoidable.
    
    Consequently, the Stokes vectors, $\vec{S}_1(\theta)$ and $\vec{S}_2(\theta)$, of the first two scattering orders were calculated for one hemisphere of a planet with a homogeneous, semi-infinite atmosphere using a plane-parallel approximation in App. \ref{app:infinitely_optically_thick} (see references therein for studies with similar calculations).
    $\vec{S}_2$ is of the form
    \begin{equation}
        \vec{S}_\mathrm{2}(\theta)
        = \int_0^{2\pi}\int_0^{\pi}
         {\vec{S}_\mathrm{out}}(\theta,\theta_1,\phi_1) \, P(\theta,\theta_1,\phi_1)
         \sin\theta_1\,\mathrm{d}\theta_1\,\mathrm{d}\phi_1,
         \label{eq:double_scattered_stokes_main}
    \end{equation}
    where $P(\theta,\theta_1,\phi_1)$ is given in Eq. \eqref{eq:P_definition}. 
    For a discussion of the relevant anti-symmetric part of $P$ with respect to the angle $\phi_1$, 
    $
        P'(\theta,\theta_1,\phi_1) = P(\theta,\theta_1,\phi_1) - P(\theta,\theta_1,2\pi - \phi_1),
    $
    see App. \ref{app:probability_function}.
    The circularly polarized component $V_2$ is 
    \begin{equation}
        V_2(\theta) = \int_0^{\pi}\int_0^\pi  V_\mathrm{out}(\theta,\theta_1,\phi_1) \, P^\prime(\theta,\theta_1,\phi_1) \sin\theta_1\,\mathrm{d}\theta_1\,\mathrm{d}\phi_1,
        \label{eq:V2_definition_main}
    \end{equation}
    where $V_\mathrm{out}$ is the circularly polarized Stokes parameter in $\vec{S}_\mathrm{out}$. The quantity
    \begin{equation}
        V^\prime(\theta,\theta_1) = \int_0^{\pi} V_\mathrm{out}(\theta,\theta_1,\phi_1)\, P^\prime(\theta,\theta_1,\phi_1) \sin\theta_1\,\mathrm{d}\phi_1
        \label{eq:V_prime_definition}
    \end{equation}
    defines the contribution of different first scattering directions to the circularly polarized flux of the second scattering order. Thus, $V^\prime$ characterizes the influence of the scattering matrix on the planetary circular polarization.

    If the atmosphere is composed of more than one particle species,
    scattering is described by an effective scattering matrix, $\mathbf{F}_\mathrm{eff}$, and an effective single scattering albedo, $\varpi_\mathrm{eff}$ \citep[see][]{martin1978,wolf2003,steinacker2013},
    \begin{equation}
        \varpi_\mathrm{eff} \mathbf{F}_\mathrm{eff} = \sum_{i} f_i\varpi_i\mathbf{F}_i, \quad 0 \leq f_i \leq 1.
        \label{eq:F_eff}
    \end{equation}
    Here, $\varpi_i$, $\mathbf{F}_i$, and $f_i$ are the single scattering albedo, scattering matrix, and relative contribution to the optical depth of the particle species $i$, respectively. $\varpi_\mathrm{eff}$ was determined through the normalization of $\mathbf{F}_\mathrm{eff}$. Inserting $\varpi_\mathrm{eff}\mathbf{F}_\mathrm{eff}$ into Eq. \eqref{eq:S_out_main} results in
    \begin{align}
        \vec{S}_{ij}(\theta,\theta_1,\phi_1) & = \frac{\varpi_i\varpi_j}{16\pi^2}\mathbf{L}(\beta)\cdot\mathbf{F}_j(\theta_2)\cdot\mathbf{L}(\phi_2)\cdot \mathbf{F}_i(\theta_1)\cdot\mathbf{L}(\phi_1) \cdot \vec{S}_\mathrm{in}, \notag \\
        \vec{S}_\mathrm{out}(\theta,\theta_1,\phi_1) & = \sum_{i}\sum_{j} f_if_j\vec{S}_{ij}(\theta,\theta_1,\phi_1),
        \label{eq:S_ij_definition}
    \end{align}
    where $f_if_j\vec{S}_{ij}$ is the contribution of scattering first by a particle of species $i$ and second by one of species $j$. Inserting this result in Eq. \eqref{eq:double_scattered_stokes_main}, we have
    \begin{align}
        \vec{S}_2(\theta) & = \sum_{i} \sum_{j} f_i f_j \vec{S}_{2,ij}(\theta), \notag \\
        \vec{S}_{2,ij}(\theta) & = \int_0^{\pi}\int_0^{\pi} \vec{S}_{ij}(\theta,\theta_1,\phi_1)\, P^\prime(\theta,\theta_1,\phi_1)\sin\theta_1\,\mathrm{d}\theta_1\,\mathrm{d}\phi_1,
        \label{eq:S_2_ij}
    \end{align}
    and, similar to Eq. \eqref{eq:V2_definition_main} and \eqref{eq:V_prime_definition},
    \begin{equation}
        V^\prime_{ij}(\theta,\theta_1) = \int_0^{\pi} V_{ij}(\theta,\theta_1,\phi_1)\, P^\prime(\theta,\theta_1,\phi_1) \sin\theta_1\,\mathrm{d}\phi_1.
        \label{eq:Vij_prime}
    \end{equation}
    
    In the following sections, we analyze the influence of different cloud particle species on the circular polarization of radiation that was scattered twice in a homogeneous planetary atmosphere.
    Because $F_{34}$ is zero for Rayleigh scattering, only interactions of linearly polarized radiation with cloud particles cause circular polarization in our framework. 
    Therefore, the two contributing processes are scattering twice by cloud particles (Sect. \ref{sec:mie_mie_scattering}) and scattering once by $\mathrm{H_2}$ molecules before scattering by cloud particles (Sect. \ref{sec:rayleigh_mie_contribution}).

\subsection{Contribution of double Mie scattering}
\label{sec:mie_mie_scattering}

    \begin{figure*}
        \centering
        \includegraphics{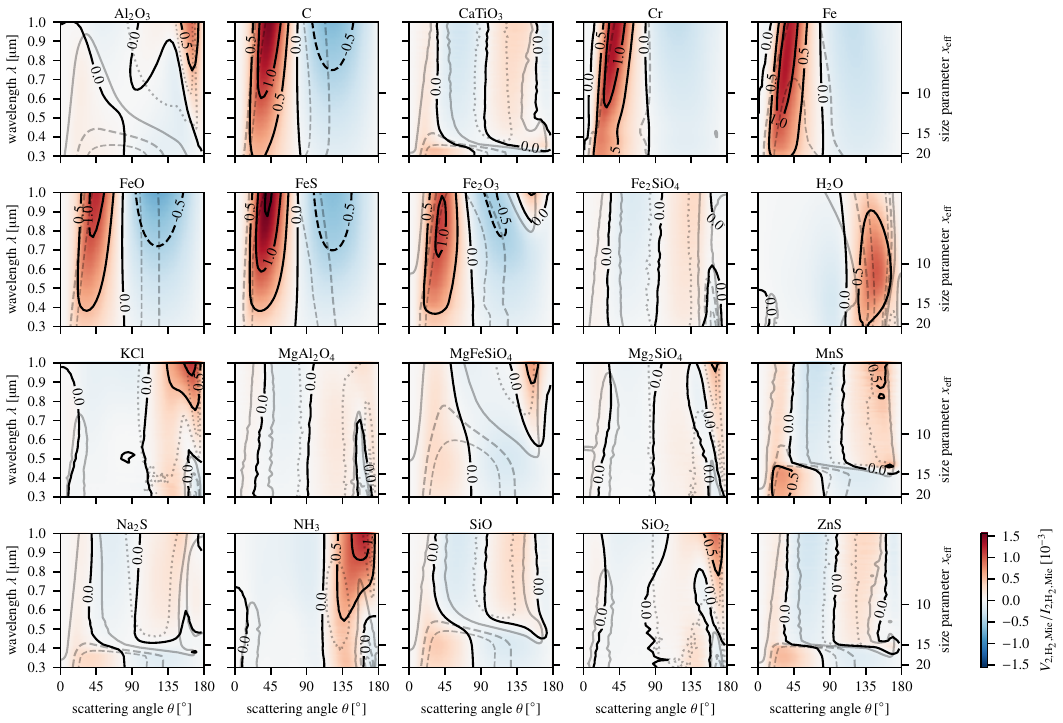}
        \caption{Circular polarization, $P_\mathrm{c2,Mie,Mie}$, of the contribution, $\vec{S}_\mathrm{2,Mie,Mie}$, to the second scattering order, $\vec{S}_2$, of a planetary hemisphere with a homogeneous atmosphere in units of $10^{-3}$ for various cloud condensates and wavelengths visualized using a color map with black contour lines.  For $\theta > 180\degr$, $P_\mathrm{c}(\theta) = -P_\mathrm{c}(2\pi-\theta)$. Overlaid gray contour lines visualize the signed degree of single scattering linear polarization, $P_\mathrm{s}$, from \citet{lietzow2022}. Solid gray lines mark $P_\mathrm{s} = 0$, dotted lines mark $-0.2$ (thick) and $-0.4$ (thin), and dashed lines mark $0.2$ (thick) and $0.4$ (thin).}
        \label{fig:mie_mie_scattering}
    \end{figure*}

    $
        P_\mathrm{c2,Mie,Mie} = {V_\mathrm{2,Mie,Mie}}/{I_\mathrm{2,Mie,Mie}}, 
    $  
    the degree of circular polarization of radiation that was scattered twice by cloud particles,
    is visualized for wavelengths between $0.3\ \mathrm{\umu m}$ and $1.0\ \mathrm{\umu m}$ in Fig.~\ref{fig:mie_mie_scattering}. The sign of $P_\mathrm{c2,Mie,Mie}$ in Fig.~\ref{fig:mie_mie_scattering} is correct for the upper hemisphere of a planet if the illuminated crescent is to the right. For comparison, overlaid contour lines visualize the signed degree of linear polarization of the first scattering order, $P_\mathrm{s} = -F_{12}/F_{11}$ \citep[see also][]{lietzow2022}. 
    The contribution, $V^\prime_\mathrm{Mie,Mie}(\theta,\theta_1)$, of first scattering angles, $\theta_1$, to $V_\mathrm{2,Mie,Mie}$ is visualized in Fig.~\ref{fig:mie_mie_features_500nm} and \ref{fig:mie_mie_features_700nm} for wavelengths of $0.5\ \mathrm{\umu m}$ and $0.7\ \mathrm{\umu m}$, respectively. The behavior of $V^\prime_\mathrm{Mie,Mie}$ reveals two major contributions to the circularly polarized flux of the second scattering order.
    \begin{enumerate}[label = (\roman*)]
        \item \label{item:forward_process} In the first process, radiation is scattered in directions in the vicinity of the direction toward the observer first, such that $\theta_1$ is close to $\theta$. Thereafter, it is scattered forward such that the second scattering angle, $\theta_2$, is small. This contribution is connected to a positive peak in $F_{34}/F_{11}$ at small scattering angles (see a in Fig. \ref{fig:F34}).
        When this process dominates, such as for $\mathrm{MgFeSiO_4}$ (see Fig. \ref{fig:mie_mie_scattering}, \ref{fig:mie_mie_features_500nm}, and \ref{fig:mie_mie_features_700nm}), changes in the handedness of $P_\mathrm{c2,Mie,Mie}(\theta)$ typically occur between two sign changes of $P_\mathrm{s}(\theta)$ in the vicinity of extrema of the single scattering linear polarization as described in App. \ref{app:process_i}.
        
        \item \label{item:backward_process} In the second process, radiation is scattered in directions that lie in the vicinity of the direction pointing away from the observer first, such that $\theta_1 \approx 180\degr - \theta$, and then backward toward the observer. In this case, $ 180\degr - \theta_2$ is small. Such contributions are connected to the negative peak (b) in $F_{34}/F_{11}$ if (b) is located at scattering angles larger than $120\degr$ (see Fig. \ref{fig:F34}), or to the positive peak (c).
        If this process dominates, such as for $\mathrm{SiO_2}$ (see Fig. \ref{fig:mie_mie_scattering}, \ref{fig:mie_mie_features_500nm}, and \ref{fig:mie_mie_features_700nm}), zeros of $P_\mathrm{c2,Mie,Mie}(\theta)$ are located between zeros of $P_\mathrm{s}(180\degr-\theta)$ instead.
    \end{enumerate}
    For Rayleigh-Mie scattering, the two processes are, in principle, the basis of the model given by \citet{wolstencroft1976}.
    Often, one change in the handedness of $P_\mathrm{c2,Mie,Mie}$ occurs at an overall scattering angle, $\theta_\mathrm{geom}$, close to $90\degr$, similar to the geometric effect suggested by \citet{kawata1978}.
    Given that circular polarization is induced when linearly polarized light is scattered by cloud particles, \citet{kawata1978} argued that a resemblance of features in $P_\mathrm{s}$ and $P_\mathrm{c2,Mie,Mie}$ is expected, implying that $P_\mathrm{c2,Mie,Mie}$ is dominated by process \ref{item:forward_process}. While this is a major contribution for selected materials, other contributions are not negligible.

\subsubsection{Large imaginary part of the refractive index}
\label{sec:high_k}

    At a wavelength of $0.5\ \mathrm{\umu m}$, $\mathrm{FeO}$ has an imaginary part, $k$, of the refractive index of $0.74$, $\mathrm{Fe_2O_3}$ of $0.78$, graphite of $1.2$, $\mathrm{FeS}$ of $1.4$, iron of $3.1$, and chromium of $4.5$.
    For all of these materials, the second scattering induces the largest circular polarization when linearly polarized radiation is scattered by a small angle, $\theta_2$, between $5\degr$ and $30\degr$ due to a positive peak in $F_{34}/F_{11}$ (see feature a in Fig. \ref{fig:F34}).
    Therefore, the main contributions to $P_\mathrm{c2,Mie,Mie}$ for these materials are from process \ref{item:forward_process} (see Fig. \ref{fig:mie_mie_features_500nm} and \ref{fig:mie_mie_features_700nm}).
    
    Excluding $\mathrm{Fe_2O_3}$ at large wavelengths, $P_\mathrm{s}$ is positive at all scattering angles and considered wavelengths for the aforementioned particle species \citep{lietzow2022}. Thus, the handedness of $P_\mathrm{c2,Mie,Mie}$ changes once at an overall scattering angle, $\theta_\mathrm{geom}$, of approximately $80\degr$. 
    For overall scattering angles, $\theta$, smaller than $\theta_\mathrm{geom}$, $P_\mathrm{c2,Mie,Mie}$ is positive, while negative values occur when $\theta$ is larger than $\theta_\mathrm{geom}$ (see Fig. \ref{fig:mie_mie_scattering}). The positive branch of $P_\mathrm{c2,Mie,Mie}$ reaches higher absolute values, as $P_\mathrm{s}$ is larger at scattering angles smaller than $\theta_\mathrm{geom}$ \citep[see][]{lietzow2022}. The maximum of $P_\mathrm{c2,Mie,Mie}$ is between $20\degr$ and $45\degr$ and moves to larger $\theta$ for larger wavelengths.     
    
    For chromium, iron, and $\mathrm{FeS}$ particles, additional contributions are found in $V^\prime_\mathrm{Mie,Mie}$ when $\theta$ is larger than $90\degr$. A positively circularly polarized feature of $V^\prime_\mathrm{Mie,Mie}$ appears at $\theta_1$ between $90\degr$ and $135\degr$, while a negatively circularly polarized contribution is found at $\theta_1$ smaller than $90\degr$. The influence of these features decreases toward larger wavelengths (comp. Fig. \ref{fig:mie_mie_features_500nm} and \ref{fig:mie_mie_features_700nm}). 
    Because $P_\mathrm{s}$ is larger at scattering angles below $90\degr$, the negatively circularly polarized feature in $V^\prime_\mathrm{Mie,Mie}$ dominates. To determine second scattering angles, $\theta_2$ (see App. \ref{app:second_scattering_angles}), that correspond to these features, it is necessary to investigate $V_\mathrm{Mie,Mie} P^\prime$ (see Eq. \ref{eq:Vij_prime}). This is demonstrated for iron in App. \ref{app:iron}. The positively circularly polarized feature in $V^\prime_\mathrm{Mie,Mie}$ corresponds to $\theta_2$ between $30\degr$ and $90\degr$, and thus to scattering in directions corresponding to the negative peak (b) in $F_{34}/F_{11}$. Meanwhile, the negative feature in $V^\prime_\mathrm{Mie,Mie}$ corresponds to $\theta_2$ larger than $90\degr$, and thus to the shallower parts of the negative plateau in $F_{34}/F_{11}$ (see Fig. \ref{fig:F34}). 
    While different radiation paths contribute to the circularly polarized flux for chromium, iron, and $\mathrm{FeS}$, the resulting $P_\mathrm{c2,Mie,Mie}$ behaves similarly as in the case of graphite, $\mathrm{FeO}$, and $\mathrm{Fe_2O_3}$ (see Fig. \ref{fig:mie_mie_scattering}).
    
    Deviating from the general behavior, $P_\mathrm{c2,Mie,Mie}$ is negative for chromium for $\theta \lesssim 5\degr$ and wavelengths greater than $0.4\ \mathrm{\umu m}$. For larger wavelengths, the feature extends to larger $\theta$, reaching $14\degr$ at $1.0\ \mathrm{\umu m}$. A similar feature is found for iron above $0.57\ \mathrm{\umu m}$ (see Fig. \ref{fig:mie_mie_scattering}). The wavelength dependence correlates with the amplitude of the positive peak (a) in $F_{34}/F_{11}$ at small scattering angles (comp. Fig. \ref{fig:F34}).
    
    For $\mathrm{Fe_2O_3}$, $k$ decreases with increasing wavelength. At wavelengths above $0.84\ \mathrm{\umu m}$, a negative plateau of $P_\mathrm{s}$ is found at scattering angles around $150\degr$ \citep{lietzow2022}. Through process \ref{item:forward_process}, this linear polarization feature induces a positive plateau of $P_\mathrm{c2,Mie,Mie}$ at overall scattering angles, $\theta$, between $156\degr$ and $160\degr$. The plateau spans a wider range of scattering angles at longer wavelengths. At $1.0\ \mathrm{\umu m}$, it extends from $135\degr$ to $173\degr$.
    $\mathrm{MgFeSiO_4}$ particles show a similar behavior to $\mathrm{Fe_2O_3}$ particles, with a plateau of positive $P_\mathrm{c2,Mie,Mie}$ appearing at wavelengths larger than $0.57\ \mathrm{\umu m}$ (see Fig. \ref{fig:mie_mie_scattering}).

\subsubsection{Influence of the real part of the refractive index}
\label{sec:diverse_n}

    For condensates with a complex part, $k$, of the refractive index below $0.01$, the behavior of $P_\mathrm{c2,Mie,Mie}$ is determined by the real part, $n$, of their refractive index. This includes water ice ($n = 1.31$ at $0.5\ \mathrm{\umu m}$), ammonia ice ($1.44$), $\mathrm{KCl}$ ($1.50$), $\mathrm{SiO_2}$ ($1.55$), $\mathrm{Mg_2SiO_4}$ ($1.62$), $\mathrm{MgAl_2O_4}$ ($1.72$), and $\mathrm{Fe_2SiO_4}$ ($1.85$) particles.
    For all seven materials, large absolute values of $F_{34}/F_{11}$ are found in a negative plateau at scattering angles of between $130\degr$ and $170\degr$ (see feature b in Fig. \ref{fig:F34}). Thus, process \ref{item:backward_process} is equally important as process \ref{item:forward_process} for these materials. Only for the materials with the largest $n$, $\mathrm{MgAl_2O_4}$ and $\mathrm{Fe_2SiO_4}$, is process \ref{item:forward_process} dominant. Process~\ref{item:backward_process} contributes less for $\mathrm{MgAl_2O_4}$ and $\mathrm{Fe_2SiO_4}$, because the influences of the negative peak (b) and the positive peak (c) in $F_{34}/F_{11}$ (see Fig. \ref{fig:F34}) tend to cancel out.
    
    When $k$ is small, the single scattering linear polarization, $P_\mathrm{s}$, is predominantly negative \citep{lietzow2022}, which results in an inverse general trend in the circular polarization compared to materials with large $k$. $P_\mathrm{c2,Mie,Mie}$ is predominantly negative when $\theta$ is smaller than $\theta_\mathrm{geom}$ and positive if $\theta$ is larger than $\theta_\mathrm{geom}$ (see Fig. \ref{fig:mie_mie_scattering}).
    Water ice, ammonia ice, and $\mathrm{KCl}$ have $n$ of at most $1.5$ and $\theta_\mathrm{geom}$ around $110\degr$. $\mathrm{Mg_2SiO_4}$, $\mathrm{MgAl_2O_4}$, and $\mathrm{Fe_2SiO_4}$ have $n$ larger than $1.6$ and $\theta_\mathrm{geom}$ close to $90\degr$. Between real parts of the refractive index of $1.5$ and $1.6$, a transition between these behaviors occurs. For $\mathrm{SiO_2}$, $n$ is between $1.58$ and $1.54$ at the considered wavelengths. Thus, $\theta_\mathrm{geom}$ is wavelength-dependent for $\mathrm{SiO}_2$ with values between $75\degr$ and $105\degr$.
    Depending on the material and the wavelength, two additional features of $P_\mathrm{c2,Mie,Mie}$ occur, which break the general trend.

    First, a negative plateau of $P_\mathrm{c2,Mie,Mie}$ occurs at large $\theta$ for all materials except for water ice. For most materials, this feature is related to a positive plateau of $P_\mathrm{s}$ at small scattering angles due to the large influence of process \ref{item:backward_process}. In some cases, an additional contribution to the feature is from process \ref{item:forward_process} due to the positive plateau of $P_\mathrm{s}$ at large scattering angles related to the primary rainbow \citep[see e.g.,][]{liou1971,hansen1974b}.

    For water ice, the negative plateau of $P_\mathrm{c2,Mie,Mie}$ at large scattering angles is missing.
    As seen in Fig. \ref{fig:mie_mie_features_500nm} and \ref{fig:mie_mie_features_700nm}, the primary rainbow in $P_\mathrm{s}$ leads to a negative contribution to $P_\mathrm{c2,Mie,Mie}$ from process \ref{item:forward_process} for water ice. However, this effect is overpowered by a large positive contribution at overall scattering angles around $150\degr$ from process \ref{item:backward_process}.
    
    For ammonia ice particles, a negative plateau of $P_\mathrm{c2,Mie,Mie}$ is located at $\theta$ larger than $161\degr$ and wavelengths up to $0.56\ \mathrm{\umu m}$ (see Fig.~\ref{fig:mie_mie_scattering}). For $\mathrm{KCl}$ particles, it is located at wavelengths up to $0.54\ \mathrm{\umu m}$. As the primary rainbow is only found up to a wavelength of $0.5\ \mathrm{\umu m}$ for $\mathrm{KCl}$ \citep[see][]{lietzow2022}, the negative contribution of process \ref{item:backward_process} is overpowered by the positive contribution of process \ref{item:forward_process} for wavelengths above $0.54\ \mathrm{\umu m}$ (see Fig. \ref{fig:mie_mie_features_700nm}). 
    For $\mathrm{SiO_2}$ particles, the negative plateau in $P_\mathrm{c2,Mie,Mie}$ at large $\theta$ extends to wavelengths of up to $0.81\ \mathrm{\umu m}$. For increasing wavelengths, it moves to smaller $\theta$, such that the sign change occurs at a $\theta$ of $158\degr$ at $0.3\ \mathrm{\umu m}$ and at $138\degr$ at $0.7\ \mathrm{\umu m}$ (see Fig. \ref{fig:mie_mie_scattering}).
    At a wavelength of $0.41\ \mathrm{\umu m}$, $P_\mathrm{c2,Mie,Mie}$ is positive again at $\theta\gtrsim 175\degr$ for $\mathrm{SiO_2}$. For larger wavelengths, the plateau of negative $P_\mathrm{c2,Mie,Mie}$ ends at smaller $\theta$, ending at $150\degr$ for $0.8\ \mathrm{\umu m}$ (see Fig. \ref{fig:mie_mie_scattering}).
    Similarly for $\mathrm{MgSiO_4}$, the negative plateau in $P_\mathrm{c2,Mie,Mie}$ at large $\theta$ is located between $156\degr$ and $175\degr$ at $0.3\ \mathrm{\umu m}$. It is observed across the entire discussed wavelength range and shifts to smaller $\theta$ as the wavelength increases. At $1.0\ \mathrm{\umu m}$, it is located at $\theta$ between $133\degr$ and $141\degr$ (see Fig. \ref{fig:mie_mie_scattering}).

    For $\mathrm{MgAl_2O_4}$ and $\mathrm{FeSiO_4}$, the negative plateau of $P_\mathrm{c2,Mie,Mie}$ at large $\theta$ is instead related to a positive plateau of $P_\mathrm{s}$ at large scattering angles, as process \ref{item:forward_process} dominates.
    Consequently, the negative plateau at large $\theta$ appears at similar wavelengths as the positive plateau in $P_\mathrm{s}$ \citep[see][]{lietzow2022} for $\mathrm{MgAl_2O_4}$. For $\mathrm{Fe_2SiO_4}$ particles, the negative plateau in $P_\mathrm{s}$ at large $\theta$ covers a small range of scattering angles at wavelengths between $0.64\ \mathrm{\umu m}$ and $0.85\ \mathrm{\umu m}$ \citep{lietzow2022}, such that no positive plateau in $P_\mathrm{c2,Mie,Mie}$ occurs (see App. \ref{app:process_i}).
    
    Second, a positive plateau of $P_\mathrm{c2,Mie,Mie}$ occurs at small $\theta$ for all materials, but at different material-dependent wavelengths and overall scattering angles. This feature is caused by positive plateaus of $P_\mathrm{s}$ at small scattering angles through process \ref{item:forward_process}, which dominates for most materials at small $\theta$. In some cases, process \ref{item:backward_process} strengthens the feature through contributions related to the primary rainbow feature.
    
    For $\mathrm{H_2O}$ ice, a positive plateau in $P_\mathrm{s}$ at wavelengths shorter than $0.5\ \mathrm{\umu m}$ at scattering angles around $15\degr$ \citep[see][]{lietzow2022} causes a positive plateau in $P_\mathrm{c2,Mie,Mie}$ at $\theta$ of less than $15\degr$ (see Fig. \ref{fig:mie_mie_scattering}).
    For $\mathrm{NH_3}$ ice, a positive plateau in $P_\mathrm{c2,Mie,Mie}$ is found at $\theta$ of less than $15\degr$ up to a wavelength of $0.72\ \mathrm{\umu m}$ (see Fig. \ref{fig:mie_mie_scattering}). A plateau of positive $P_\mathrm{s}$ at similar scattering angles is found up to $0.65\ \mathrm{\umu m}$ \citep{lietzow2022}. For wavelengths above $0.65\ \mathrm{\umu m}$, negative contributions of process~\ref{item:backward_process} strengthen the feature (see Fig. \ref{fig:mie_mie_features_700nm}). 
    Similarly, a positive plateau of $P_\mathrm{c2,Mie,Mie}$ at $\theta$ smaller than $20\degr$ appears in the entire wavelength range for $\mathrm{KCl}$, $\mathrm{SiO_2}$, and $\mathrm{MgSiO_4}$, even though the positive plateau of $P_\mathrm{s}$ is only found up to wavelengths of between $0.75\ \mathrm{\umu m}$, $0.8\ \mathrm{\umu m}$, and $0.95\ \mathrm{\umu m}$, respectively \citep[see][]{lietzow2022}, due to the influence of process \ref{item:backward_process}.
    For $\mathrm{MgAl_2O_4}$ and $\mathrm{Fe_2SiO_4}$ particles, the positive plateau in $P_\mathrm{c2,Mie,Mie}$ at $\theta$ below $30\degr$ is found in the entire wavelength range, similar to the positive plateau of $P_\mathrm{s}$, as expected due to the dominance of process \ref{item:forward_process}.
    
\subsubsection{Variable imaginary part of the refractive index}
\label{sec:variable_k}

    Our optical dataset includes materials for which $k$ decreases with increasing wavelength \citep[and refs. in Table \ref{tab:particle_species}]{kitzmann2018}. Out of these, $\mathrm{Fe_2O_3}$ and olivine were discussed in Sect. \ref{sec:high_k}. Decreases in $k$ are also found for $\mathrm{Al_2O_3}$, $\mathrm{CaTiO_3}$, $\mathrm{MnS}$, $\mathrm{NaCl}$, $\mathrm{SiO}$, and $\mathrm{ZnS}$.
    At a wavelength of $0.3\ \mathrm{\umu m}$, the behavior of $P_\mathrm{c2,Mie,Mie}$ is similar to particle species with high $k$ discussed in Sect. \ref{sec:high_k}. This behavior extends up to a $\theta$ and material-dependent wavelength $\lambda_k$. At $\lambda_k$, a transition occurs in the behavior of $P_\mathrm{c2,Mie,Mie}$ and $P_\mathrm{s}$ \citep[see][]{lietzow2022}. For wavelengths larger than $\lambda_k$, the behavior is more similar to that of materials in Sect. \ref{sec:diverse_n}. 
    
    For $\mathrm{CaTiO_3}$, $\mathrm{SiO}$, and $\mathrm{ZnS}$, $n$ is larger than $2$ and process~\ref{item:forward_process} dominates (see Fig. \ref{fig:mie_mie_features_500nm} and \ref{fig:mie_mie_features_700nm}). For these materials, the negative plateau in $P_\mathrm{c2,Mie,Mie}$ at large overall scattering angles extends to an overall scattering angle, $\theta$, of $180\degr$ at wavelengths larger than $\lambda_k$, similar to the positive plateau in $P_\mathrm{s}$ \citep[see][]{lietzow2022}. For $\mathrm{MnS}$, the negative plateau in $P_\mathrm{c2,Mie,Mie}$ at large $\theta$ is not found. The positive plateau in $P_\mathrm{s}$ occurs only at scattering angles larger than $165\degr$ \citep{lietzow2022} and, consequently, no negative plateau in $P_\mathrm{c2,Mie,Mie}$ is found (comp. App. \ref{app:process_i}).

\subsection{Contribution of Rayleigh-Mie scattering}
\label{sec:rayleigh_mie_contribution}

    \begin{figure*}
        \centering
        \includegraphics{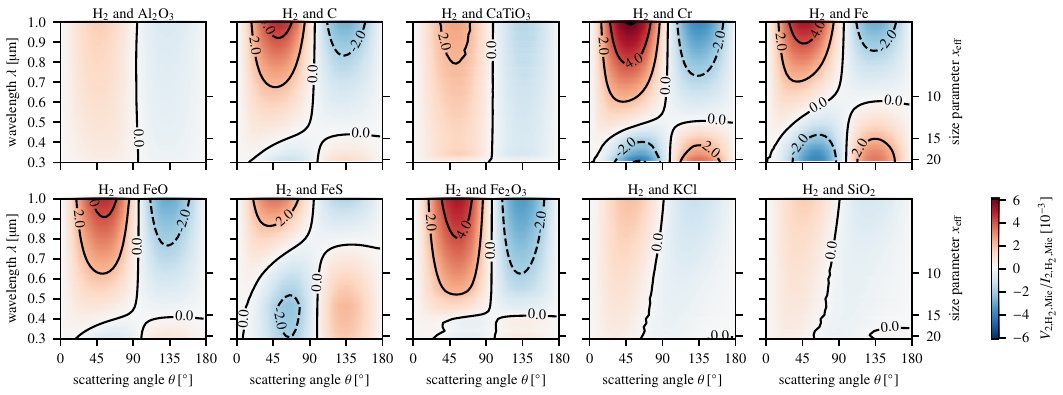}
        \caption{Circular polarization $P_\mathrm{c2,H_2,Mie}$ of the contribution $\vec{S}_\mathrm{2,H_2,Mie}$ of scattering once by $\mathrm{H_2}$ molecules before scattering once by Mie scattering cloud particles to the Stokes vector of the second scattering order, $\vec{S}_2$, for various cloud condensates and wavelengths visualized using a color map with black contour lines.}
        \label{fig:rayleigh_mie_scattering}
    \end{figure*}
    
    The second contribution to $V_2$ is from radiation that is scattered once by $\mathrm{H_2}$ molecules before cloud particles scatter it toward the observer. Results for the degree of circular polarization
    $
        P_\mathrm{c2,H_2,Mie} = {V_\mathrm{2,H_2,Mie}}/{I_\mathrm{2,H_2,Mie}}
    $
    for mixtures of $\mathrm{H_2}$ molecules with one species of cloud particles are visualized for wavelengths of between $0.3\ \mathrm{\umu m}$ and $1.0\ \mathrm{\umu m}$ in Fig. \ref{fig:rayleigh_mie_scattering} and \ref{fig:mixed_scattering_2}. The contribution $V^\prime_\mathrm{H_2,Mie}(\theta,\theta_1)$ of different first scattering angles, $\theta_1$, is visualized in Fig. \ref{fig:H2_Mie_features} for a wavelength of $0.5\ \mathrm{\umu m}$. 
    
    As in double Mie scattering, the largest contributions are from processes~\ref{item:forward_process} and \ref{item:backward_process} for most materials. In that case, $P_\mathrm{c2,H_2,Mie}$ has a similar behavior to $\mathrm{Al_2O_3}$ or $\mathrm{CaTiO_3}$ particles, which are showcased in Fig. \ref{fig:rayleigh_mie_scattering}. For all wavelengths, a single change in the sign of $P_\mathrm{c2,H_2,Mie}$ occurs at a $\theta_\mathrm{geom}$ between $90\degr$ and $100\degr$, as the single scattering linear polarization of $\mathrm{H_2}$ molecules is positive at all scattering angles \citep[e.g.,][]{hansen1974b}. $P_\mathrm{c2,H_2,Mie}$ is positive for $\theta$ that are larger than $\theta_\mathrm{geom}$ and negative for $\theta$ of less than $\theta_\mathrm{geom}$. Notably, these results are in agreement with those by \citet{wolstencroft1976} for scattering by $\mathrm{H_2}$ molecules and $\mathrm{NH_3}$ aerosols.

    A different behavior occurs if radiation paths not included in processes~\ref{item:forward_process} and \ref{item:backward_process} contribute significantly.
    For $\mathrm{KCl}$ and $\mathrm{SiO_2}$ particles, $\theta_\mathrm{geom}$ increases with the wavelength from $60\degr$ at $0.3\ \mathrm{\umu m}$ to $90\degr$ at $1.0\ \mathrm{\umu m}$. For wavelengths shorter than $0.35\ \mathrm{\umu m}$, a positive plateau in $P_\mathrm{c2,H_2,Mie}$ is found for $\mathrm{SiO_2}$ at $\theta$ larger than $135\degr$. For $\mathrm{KCl}$, a similar plateau occurs for wavelengths shorter than $0.3\ \mathrm{\umu m}$ (see Fig. \ref{fig:rayleigh_mie_scattering}). The main contribution to this feature is from paths with $\theta_1$ between $100\degr$ and $130\degr$ and $\theta_2$ around $60\degr$. A similar behavior to $\mathrm{KCl}$ is found for $\mathrm{Mg_2SiO_4}$ (Fig. \ref{fig:mixed_scattering_2}).

    For cloud particles with large $k$, a change in the handedness of $P_\mathrm{c2,H_2,Mie}$ occurs at a $\theta_\mathrm{geom}$ around $90\degr$. When $\theta$ is smaller than $\theta_\mathrm{geom}$, $P_\mathrm{c2,H_2,Mie}$ is negative for shorter wavelengths and positive for larger wavelengths (see Fig. \ref{fig:rayleigh_mie_scattering}). The transition occurs at a scattering-angle-dependent wavelength, $\lambda_g$. 
    For materials with large $k$, the two important features of $F_{34}/F_{11}$ are a positive peak (a) at scattering angles between $5\degr$ and $30\degr$, which moves to larger scattering angles for larger wavelengths, and a negative peak (b) at scattering angles around $60\degr$ (see Fig. \ref{fig:F34}). 
    $\lambda_g$ reflects which of the two features is more relevant at a given $\theta$ and wavelength. If the positive peak (a) is more relevant, process \ref{item:forward_process} dominates. Otherwise, the influence of paths with $\theta_2$ around $60\degr$ is more important, which results in a negative $P_\mathrm{c2,H_2,Mie}$ at $\theta$ smaller than $\theta_\mathrm{geom}$. For small $\theta$, $\theta_2$ must also be small. This limits the influence of the negative peak (b) at small $\theta$. With increasing $\theta$ from $0\degr$ to $90\degr$, $\lambda_g$ therefore increases. For $\theta$ larger than $90\degr$, $\lambda_g$ remains constant or decreases slightly.
    
    The transition is fully visible for chromium and iron in Fig. \ref{fig:rayleigh_mie_scattering}. For $\theta$ larger than $110\degr$, $\lambda_g$ is around $0.5\ \mathrm{\umu m}$ for chromium and around $0.6\ \mathrm{\umu m}$ for iron.
    For graphite, $\lambda_g$ is smaller than the shortest considered wavelength of $0.3\ \mathrm{\umu m}$ for $\theta$ of less than $15\degr$. When $\theta$ is larger than $110\degr$, $\lambda_g$ reaches at most $0.44\ \mathrm{\umu m}$. A very similar behavior is found for $\mathrm{FeO}$ and $\mathrm{Fe_2O_3}$, for which $\lambda_g$ is around $0.4\ \mathrm{\umu m}$ for large $\theta$.
    For $\mathrm{FeS}$, $P_\mathrm{c2,H_2,Mie}$ is positive at $\theta$ of less than $24\degr$ for a wavelength of $0.3\ \mathrm{\umu m}$. $\lambda_g$ rises steeply with $\theta$. When $\theta$ is larger than $110\degr$, $\lambda_g$ reaches $0.75\ \mathrm{\umu m}$.

\section{Radiative transfer simulations}
\label{sec:results}
    \begin{table}
        \centering
        \caption{Properties of the model atmosphere.}
        \begin{tabular}{rcc}
            \hline\hline\\[-0.9em]
            & Gas & Clouds \\ 
            \hline\\ [-0.9em]
            $p_\mathrm{max}\ [\mathrm{bar}]$     & $100$ & $1$ \\ 
            $p_\mathrm{min}\ [\mathrm{bar}]$     & $10^{-5}$ & $0.1$ \\
            Optical depth & 18 & 10 \\
            Composition          & $\mathrm{H_2}$ molecules & Table \ref{tab:particle_species} \\ 
            Size distribution    & none & Eq. \eqref{eq:size_distribution} \\ 
            Effective radius    & - & $1\ \mathrm{\umu m}$ \\ 
            Effective variance  & - & $0.1$ \\
            \hline
        \end{tabular}
        \label{tab:model}
    \end{table}

    The results of Sect. \ref{sec:semi_analytical_calculations} only include the first two scattering orders. In this section, we determine whether similar features occur in a more realistic scenario for a wavelength of $0.5\ \mathrm{\umu m}$.
    For this purpose, three-dimensional Monte Carlo radiative transfer simulations with POLARIS\footnote{\url{https://github.com/polaris-MCRT/POLARIS}} \citep{reissl2016,lietzow2021} were performed to calculate the planetary Stokes vector including contributions of all scattering orders. 
    
    The atmospheric model used in the simulations (see Table \ref{tab:model}) was identical to the one used in the study by \citet{lietzow2022}.
    The atmosphere was described on a logarithmically spaced pressure grid with a top pressure of $10^{-5}\ \mathrm{bar}$ and a bottom pressure of $100\ \mathrm{bar}$. A cloud layer with homogeneous density was placed between $1\ \mathrm{bar}$ and $0.1\ \mathrm{bar}$ with an optical depth of $10$ at a wavelength of $0.5\ \mathrm{\umu m}$. 
    Note that the optical depth of the atmosphere is independent of its temperature profile in this setup, such that its choice has a negligible effect on the results.
    Additionally, the contribution of planetary thermal emission to the total flux is insignificant at a wavelength of $0.5\ \mathrm{\umu m}$.
    For all simulations, the model atmosphere was therefore isothermal. 
    Cell borders and gas number densities were calculated assuming hydrostatic equilibrium and an ideal gas
    using the formulae in \citet{lietzow2021}.
    The overall optical depth of the atmosphere was $28$ at a wavelength of $0.5\ \mathrm{\umu m}$. 
    As this would be a sufficient approximation of a semi-infinite atmosphere \citep[e.g.,][]{buenzli2009}, the properties of the lower boundary of the atmosphere did not influence the results. 
    
    \begin{figure*}
        \centering
        \includegraphics{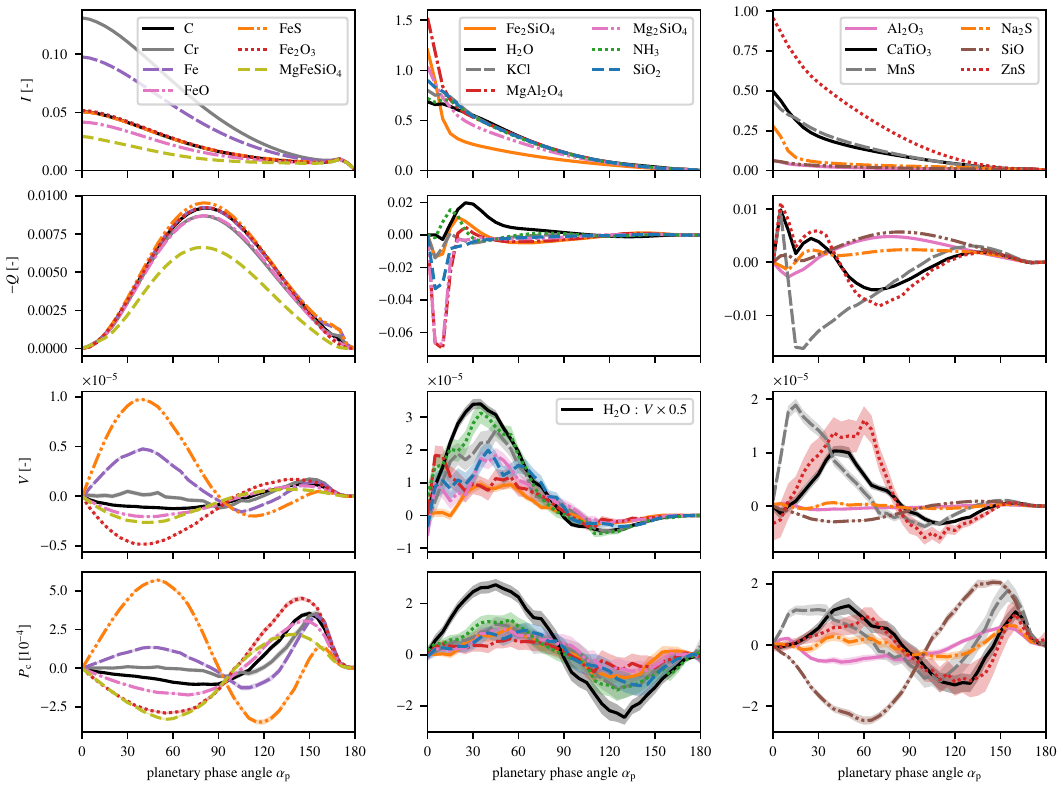}
        \caption{First row: Planetary intensity phase curve at a wavelength of $0.5\ \mathrm{\umu m}$, normalized such that $I(0\degr)$ is equal to the geometric albedo. Second row: Linearly polarized intensity, $-Q$, as a function of the planetary phase angle, $\alpha_\mathrm{p}$, at a wavelength of $0.5\ \mathrm{\umu m}$. The sign is chosen to facilitate comparisons with the signed degree of linear polarization, $-Q/I$. Third row: Circularly polarized intensity, $V$, of the hemisphere above the planetary scattering plane as a function of $\alpha_\mathrm{p}$ at a wavelength of $0.5\ \mathrm{\umu m}$. The statistical error of POLARIS simulations using $10^{10}$ photon packages is shown as an error band. For water, $V$ was multiplied by $0.5$ to facilitate the visibility of features for the other materials. Fourth row: Circular polarization, $P_\mathrm{c}$, as a function of $\alpha_\mathrm{p}$ of the hemisphere above the planetary scattering plane. Left column: Materials with a high imaginary part, $k$, of the refractive index. Central column: Materials with $k < 0.01$ and various real parts. Right column: Materials with a variable $k$ between $0.3\ \mathrm{\umu m}$ and $1.0\ \mathrm{\umu m}$.}
        \label{fig:all_materials}
    \end{figure*}

    Results are given as a function of the planetary phase angle, $\alpha_\mathrm{p}$, in $5\degr$ steps. $\alpha_\mathrm{p}$ is the angle between the directions from the planet toward the star and the observer.
    An overview of the results for the total intensity, $I$, the linearly polarized intensity, $Q$, the circularly polarized intensity, $V$, and the intrinsic degree of circular polarization, $P_\mathrm{c}$, is given in Fig. \ref{fig:all_materials}. As is common in numerical studies \citep[e.g.,][]{stam2004,lietzow2021}, the Stokes vector was normalized so that $I$ equals the geometric albedo, $A_\mathrm{g}$, of the planet at a planetary phase angle of $0\degr$. 
    While $I$ and $Q$ were integrated on the entire planet, $V$ and $P_\mathrm{c}$ were calculated for one planetary hemisphere. Realistic asymmetries such as circumplanetary rings are discussed in Sect. \ref{sec:discussion}.

    \subsection{Large imaginary part of the refractive index}
    \label{sec:sub_results_large_k}

    Results for cloud particle species with a large $k$ are presented in the left column of Fig. \ref{fig:all_materials}. 
    If scattering by cloud particles were the dominant process inducing circular polarization, negative values of $V$ would be expected for all materials with large $k$ at $\alpha_\mathrm{p}$ smaller than $90\degr$ (see Sect. \ref{sec:high_k}). 
    In the model, however, the circular polarization is dominated by processes in which radiation is scattered by $\mathrm{H_2}$ molecules before being scattered by cloud particles. Thus, $V$ and $P_\mathrm{c}$ behave similar to $P_\mathrm{c2,H_2,Mie}$ (Sect. \ref{sec:rayleigh_mie_contribution}).

    For model planets with clouds composed of $\mathrm{FeS}$ or iron, $V$ is positive at $\alpha_\mathrm{p}$ of less than $90\degr$. When $\alpha_\mathrm{p}$ is larger than $90\degr$, $V$ is negative apart from a positive plateau at large $\alpha_\mathrm{p}$ above $125\degr$ for iron clouds and $145\degr$ for $\mathrm{FeS}$ clouds. The largest absolute values of $V$ are reached at $\alpha_\mathrm{p}$ between $45\degr$ and $50\degr$. The maximum of $P_\mathrm{c}$ of $5.7\cdot 10^{-4}$ is found at a similar $\alpha_\mathrm{p}$ for $\mathrm{FeS}$. For iron, the largest $P_\mathrm{c}$ of $3.4\cdot 10^{-4}$ is found at a phase angle of $155\degr$.

    For chromium clouds, small values of $P_\mathrm{c}$ of less than $10^{-5}$ are found at $\alpha_\mathrm{p}$ below $70\degr$.
    As the transition wavelength, $\lambda_g$, found for $P_\mathrm{c2,H_2,Mie}$ in the case of $\mathrm{H_2}$-chromium mixtures is close to $0.5\ \mathrm{\umu m}$ for $\theta$ larger than $110\degr$, this is expected. The positive peak in $P_\mathrm{c}$ at $\alpha_\mathrm{p}$ of $155\degr$ reaches values of $3.5\cdot 10^{-4}$ for chromium clouds.
    
    For the remaining clouds with large $k$, $V$ is negative when $\alpha_\mathrm{p}$ is smaller than $90\degr$ and positive when $\alpha_\mathrm{p}$ is larger than $90\degr$. The largest negative values of $V$ are found for $\mathrm{Fe_2O_3}$ clouds at $\alpha_\mathrm{p}$ around $40\degr$. $\mathrm{Fe_2O_3}$ clouds also reach the highest values of $P_\mathrm{c}$ at around $145\degr$ with $4.5\cdot 10^{-4}$. 
    
    Below $90\degr$, absolute values of $P_\mathrm{c}$ of at most $3.3\cdot 10^{-4}$ are found for $\mathrm{MgFeSiO_4}$, which are larger than for $\mathrm{FeO}$ with a $P_\mathrm{c}$ of at most $1.7\cdot 10^{-4}$. Values for graphite cloud particles are even smaller at $1.0\cdot 10^{-4}$. The opposite is true for $\alpha_\mathrm{p}$ above $90\degr$. At $\alpha_\mathrm{p}$ of $145\degr$, the absolute value of $P_\mathrm{c}$ reaches $3.5\cdot 10^{-4}$ for graphite. Smaller values of $P_\mathrm{c}$ are found for $\mathrm{FeO}$ and $\mathrm{MgFeSiO_4}$ with at most $3.0\cdot 10^{-4}$ and $2.2\cdot 10^{-4}$, respectively.
    
    All seven cloud species with large $k$ share a very similar phase curve of the linearly polarized flux, $Q$, but show large variation in circular polarization. In addition, the model planets with graphite, $\mathrm{FeS}$, or $\mathrm{Fe_2O_3}$ cloud particles share a very similar geometric albedo of $0.05$. If a large-scale, time-independent asymmetry is present, for instance, a circumplanetary ring, circular polarization measurements could therefore be used for differentiating between these cloud compositions. However, this would require very precise measurements of the circular polarization of light reflected by hot exoplanets (see Sect. \ref{sec:discussion}). 
    Also, double Mie scattering would be the dominant contribution if the cloud top were at lower pressure levels, due to the smaller optical depth of the gas in and above the cloud layer. The circular polarization would more closely resemble $P_\mathrm{c2,Mie,Mie}$ in that case, resulting in a more uniform behavior for graphite, $\mathrm{FeS}$, and $\mathrm{Fe_2O_3}$.
    
    When scattering by $\mathrm{H_2}$ molecules before scattering by cloud particles dominates the circular polarization signal, additional information is retrievable from the wavelength, $\lambda_g$, at which the handedness of circular polarization changes. When the wavelength dependence of $V$ is interpreted as a dependence on the size parameter, $x_\mathrm{eff}$, of the cloud particles, measurements of $\lambda_g$ allow for $x_\mathrm{eff}$ to be determined. As the refractive index of the material is wavelength-dependent \citep[see][and ref. in Table \ref{tab:particle_species}]{bohren1983}, this study only allows for a rough estimate of the influence of $x_\mathrm{eff}$, and further research is needed to verify the claim.

    \subsection{Influence of the real part of the refractive index}
    \label{sec:sub_results_small_k}

    In the central column of Fig. \ref{fig:all_materials}, we present results for cloud materials with $k$ smaller than $0.01$ and various real parts of the refractive index, ranging from $1.31$ for water ice to $1.85$ for $\mathrm{Fe_2SiO_4}$. For models with these cloud particle species, $V$ is positive when $\alpha_\mathrm{p}$ is smaller than $90\degr$ and negative when $\alpha_\mathrm{p}$ is larger than $90\degr$. This is similar to the behavior of $P_\mathrm{c2,Mie,Mie}$ in Sect. \ref{sec:diverse_n}, meaning that scattering only at cloud particles is the dominant process of inducing circular polarization for these cloud particle species.
    
    The relative statistical error of the results is high for $V$ or $P_\mathrm{c}$ even when $10^{10}$ photon packages are used. As $P_\mathrm{c}$ is on the order of $10^{-5}$ to $10^{-4}$, this is expected. The statistical error of a Monte Carlo radiative transfer simulation is proportional to the inverse square root of the number of photon packages \citep{steinacker2013}, which is also $10^{-5}$ in this case. 
    Despite the statistical noise, it is clear that most features of $P_\mathrm{c2,Mie,Mie}$ disappear when all scattering orders contribute to the result. Most notably, no sign changes other than the geometric effect at about $90\degr$ are observed, except for a shallow positive plateau for $\mathrm{Fe_2SiO_4}$ particles at $\alpha_\mathrm{p}$ above $150\degr$ that reaches a $P_\mathrm{c}$ of roughly $10^{-5}$. The existence of an even shallower positive plateau with $P_\mathrm{c}$ of less than $0.5\cdot 10^{-5}$ at high $\alpha_\mathrm{p}$ cannot be ruled out for $\mathrm{MgAl_2O_4}$, $\mathrm{MgSiO_4}$, and $\mathrm{SiO_2}$ due to the noise of the Monte Carlo simulation. A negative plateau in $P_\mathrm{c}$, expected at small phase angles according to the interpretative model, is absent for all materials.
    Compared to the interpretative model, the sign change at $\theta_\mathrm{geom}$ in $P_\mathrm{c2,Mie,Mie}$ is found at slightly larger phase angles. The various values of $\theta_\mathrm{geom}$ found in $P_\mathrm{c2,Mie,Mie}$ are contracted to a region at phase angles, $\alpha_\mathrm{p}$, between $90\degr$ and $100\degr$ when all scattering orders are considered.
    
    Notably, the amplitude of $V$ is more than twice as high for water ice cloud particles as for all other particle species. The phase curve of $P_\mathrm{c}$ for water ice clouds in this study has an amplitude of $2.7\cdot 10^{-4}$. This is comparable to results for water clouds in the atmospheres of Earth-like exoplanets \citep[e.g.,][]{rossi2018}, despite the higher optical depth of the gaseous component. 
    Our results are comparable to those by \citet{kawata1978} for atmospheres consisting purely of Mie scattering particles with a refractive index of $1.33$, which is close to the value of $1.31$ used for water ice in this study, or a refractive index of $1.44$, which is similar to ammonia ice. For particles with a size parameter of $12.5$, which corresponds to a wavelength of $0.5\ \mathrm{\umu m}$ in our study, \citet{kawata1978} found features similar to those seen in $P_\mathrm{c2,Mie,Mie}$ in Sect. \ref{sec:diverse_n} at small and large $\alpha_\mathrm{p}$.
    This suggests that the disappearance of features in our multiple scattering results is due to the influence of $\mathrm{H_2}$ molecules. The fraction of optical depth due to $\mathrm{H_2}$ is proportional to the Rayleigh scattering cross section, and thus decreases with increasing wavelength according to $\lambda^{-4}$ \citep[see, e.g.,][]{sneep2005}. At longer wavelengths of about $0.8\ \mathrm{\umu m}$ to $1.0\ \mathrm{\umu m}$, the optical depth of the gaseous component is more than $85\%$ smaller. Therefore, features of $P_\mathrm{c2,Mie,Mie}$ may show at these wavelengths.

    Due to the uniformity of the $P_\mathrm{c}$ phase curves, circular polarization measurements are not well suited to characterizing cloud particles in this range of refractive indices. In contrast, it is possible to distinguish between these materials using linear polarimetry \citep{lietzow2021}.
    An advantage of the predictable $P_\mathrm{c}$ signal of these cloud species is that it allows for a better characterization of the asymmetry. For instance, the sign of $V$ would determine which half of the planet is covered by a circumplanetary ring, which is impossible with unresolved flux or linear polarization measurements if the ring inclination longitude is close to $0\degr$ or $90\degr$ \citep[see][]{lietzow2023,veenstra2025}. In principle, this was demonstrated with observations of Saturn \citep{smith1983}.

    \subsection{Variable imaginary part of the refractive index}
    \label{sec:sub_results_variable_k}

    Finally, the right column of Fig. \ref{fig:all_materials} presents results for cloud materials with decreasing $k$ in the wavelength range between $0.3\ \mathrm{\umu m}$ and $0.5\ \mathrm{\umu m}$. As this leads to a characteristic transition in the behavior of linear polarization that occurs at different wavelengths for different materials \citep{lietzow2021}, circular polarization measurements are unnecessary to differentiate between them. Additionally, it causes large wavelength-dependent differences in the planetary geometric albedo.
    
    $P_\mathrm{c}$ and $V$ inhabit very similar features to the results for $P_\mathrm{c2,Mie,Mie}$ for these materials. They are apparently more robust against the effects of $\mathrm{H_2}$ molecules and multiple scattering than the features found for the materials discussed in Sect. \ref{sec:sub_results_small_k}. Again, scattering only by cloud particles is the dominant process.

\section{Discussion}
\label{sec:discussion}

    \subsection{Influence of realistic asymmetries}
    \label{sec:non_zero_CP}
    
    So far, all circular polarization results have been computed for one planetary hemisphere, as $P_\mathrm{c}$ would otherwise be zero due to the polar effect. In reality, a clear phase angle dependence of circular polarization is only possible as a consequence of stable large-scale asymmetries, because the planetary disk cannot be resolved in observations of exoplanets. 
    
    For example, a large part of one hemisphere of the exoplanet could be occulted by a circumplanetary ring. For Saturn, the circular polarization is much smaller for the rings than for the visible disk at optical wavelengths \citep{swedlund1972}. However, light scattered by the ring is linearly polarized \citep{kemp1973} and may still enhance the planetary circular polarization through a second scattering in the planetary atmosphere \citep{kawata1978}. Nevertheless, the small contribution to the circularly polarized flux and the invariable viewing geometry of circumplanetary rings around exoplanets would make them excellent targets for future missions aimed at observing circular polarization.
    
    Due to the additional scattered light flux, $F_\mathrm{ring}$, of the ring, the actual degree of circular polarization is smaller than the values presented in this study. Additionally, a ring blocks only a fraction, $f_\mathrm{cover}$, of the flux of one hemisphere. The additional flux from the uncovered part further reduces the circular polarization. Given the flux of one planetary hemisphere, $F_\mathrm{hemi}$, the actual $P_\mathrm{c}$ is smaller by a factor of approximately
    \begin{equation}
        \frac{F_\mathrm{hemi}}{F_\mathrm{ring} + (2-f_\mathrm{cover})F_\mathrm{hemi}}.
    \end{equation}
    $F_\mathrm{ring}$ depends on the scattering albedo of the ring particles, $f_\mathrm{cover}$ on the extent, viewing geometry, and optical depth of the ring, and $F_\mathrm{hemi}$ on planetary properties. While this formula may serve as a fast approximation, a more thorough study using radiative transfer simulations is needed to quantify the influence of circumplanetary rings. In particular, this would allow one to account for the distribution of circularly polarized flux on one planetary hemisphere.
    
    Another option is the presence of asymmetries in the atmospheric properties between the morning and evening terminators \citep[e.g.,][]{demory2013,line2016,macDonald2017,coulombe2025}. 
    For example, a part of the planetary day side near the morning terminator might be covered by clouds. In contrast, the remainder would be cloud-free as clouds dissolve due to higher dayside temperatures \citep[e.g.,][]{garcia_munoz2015b,helling2019b,roman2021,helling2021}.
    When the inclination of the planetary orbit is not $90\degr$, the scattering plane is no longer equivalent to the orbital plane of the planet.
    In the most extreme scenario of a face-on orbit, it would be perpendicular to the orbital plane. If the asymmetry were stable over time, this would result in a nonzero $P_\mathrm{c}$, although in a reduced range of phase angles. In this scenario, the additional flux of the cloud-free part again reduces the degree of circular polarization compared to the values presented in Sect. \ref{sec:results}. The exact difference depends on the cloud coverage percentage and the viewing geometry.

    \subsection{Irregularly shaped particles}
    
    As discussed in Sect. \ref{sec:diverse_n}, the sign of the circularly polarized flux of exoplanets depends on the direction of linear polarization after single scattering. For cubic $\mathrm{KCl}$ particles, this direction is measured to be perpendicular to what is expected for spherical particles \citep{hamill2024}. In that case, the handedness of $P_\mathrm{c}$ would also be opposite to that found in our results. Due to the lack of strong scattering-angle-dependent features in the matrix elements for irregularly shaped particles \citep[see, e.g.,][]{hamill2024}, the behavior of $P_\mathrm{c}$ would otherwise be as featureless as in Sect. \ref{sec:sub_results_small_k}, with a single change in the handedness of circular polarization around $90\degr$.
    
    Because the handedness of circular polarization is connected to the direction of linear polarization, the phase curve of $P_\mathrm{c}$ remains predictable. Consequently, it is still possible to use $P_\mathrm{c}$ for characterizing the ring inclination, as described in Sect. \ref{sec:sub_results_small_k}, if the linear polarization is also measured. In addition, the form of the scattering matrix is similar to Eq. \eqref{eq:scattering_matrix} if particles are randomly aligned and the particle ensemble includes mirror images of all particles \citep{van_de_hulst1957}.

    However, if particles are not only irregularly shaped but also homochiral, the matrix element $F_{14}$ is no longer zero \citep{van_de_hulst1957,wolstencroft1976}. This eliminates the polar effect and may result in a larger intrinsic planetary circular polarization, which might even serve as a biomarker for terrestrial exoplanets \citep{sparks2009a,sparks2009b,sparks2012,sparks2021}.
    
    \subsection{Observability and star-planet interactions}
    
    Not a single cloud composition considered in this study resulted in an amplitude of $P_\mathrm{c}$ of more than $6\cdot10^{-4}$, even though the stellar flux was ignored and the Stokes vector was integrated only on one planetary hemisphere. 
    When the exoplanet and its host star are not separated in observations, the planetary contribution to the circular polarization of the entire system is at most $10^{-8}$ for all condensates in our model and at most $10^{-9}$ for $\mathrm{SiO_2}$, assuming a planetary radius of $1$ Jovian radius and a star planet distance of $0.03\ \mathrm{au}$ \citep[similar to HD 189733b, see, e.g.,][]{addison2019}. 
    Polarization levels on the order of a few times $10^{-9}$ are well below the precision of a few parts per million achievable for exoplanets with current polarimeters such as HIPPI-2 \citep{bailey2020} or POLISH-2 \citep{wiktorowicz2023}. 
    Additionally, even the disk-integrated circular polarization of the quiet Sun is on the order of $0.1\ \mathrm{ppm}$ \citep{kemp1987}.
    For more active stars, star spots cause a disk-integrated circular polarization of hundreds of parts per million \citep[see][]{elias1990,wiktorowicz2025}.
    
    While the linear polarization of HD 189733b measured by \citet{wiktorowicz2025} is well explained by small $\mathrm{SiO_2}$ particles in the atmosphere of the planet, the circular polarization of the HD 189733 system is clearly of a different origin. This supports the hypothesis by \citet{wiktorowicz2025} that the observed circular polarization is due to star spots induced by star-planet interactions.
    If circular polarization signals due to star spots, which are locked to the planetary phase angle, are ubiquitous for hot Jupiter systems, this severely limits the prospects of ever observing the degree of circular polarization of hot exoplanets. Star-planet interactions lead to a plethora of effects, including planetary material loss, which produces spatially asymmetric features such as gaseous tails \citep[e.g.,][]{matsakos2015,vidotto2025}, thereby obscuring planetary circular polarization signals.
    
    Measuring the circular polarization of exoplanets, therefore, requires technical advances in the measurement of circular polarization, very stable calibrations, very limited linear polarization cross-talk, and precise modeling of star spot circular polarization. Ideally, the planet is resolved from its host star, which will be possible for cool gas giants with future coronagraphic missions such as the Habitable Worlds Observatory, thereby minimizing stellar background signals \citep{min2025}.

\section{Conclusion}
\label{sec:conclusion}

    Using an interpretative model of a planet with a homogeneous semi-infinite atmosphere, we analyzed the correlation between the optical properties of cloud particles, the planetary geometry, and the circular polarization of radiation that was scattered twice in its atmosphere for wavelengths between $0.3$ and $1\ \mathrm{\umu m}$. 
    To compare these results to realistic circular polarization phase curves, we performed three-dimensional Monte Carlo radiative transfer simulations for a wavelength of $0.5\ \mathrm{\umu m}$. Our planetary model consisted of a hydrostatic atmosphere and a cloud layer as in the study by \citet{lietzow2022}.
    As circular polarization cancels out in observations of symmetric planets due to the polar effect \citep[e.g.,][]{hansen1971,wolstencroft1976,kawata1978}, all results were integrated solely on the planetary hemispheres above the scattering plane.
    
    Depending on the complex refractive index of the considered cloud particle species, we find characteristic features in the intrinsic degree of circular polarization, $P_\mathrm{c}$, of the hemispheres of model exoplanets as a function of the planetary phase angle, $\alpha_\mathrm{p}$. However, the largest values of $P_\mathrm{c}$ are smaller than $3\cdot 10^{-4}$ for all considered cloud condensates.
    
    The general trends of $P_\mathrm{c}$ found in our second scattering order model also appear in our multiple scattering results for all considered cloud condensates. We conclude that circular polarization of starlight reflected by clouds in exoplanetary atmospheres arises mainly from the second scattering order, as presumed by \citet{hansen1971}, among others. In our planetary model, the dominant contribution comes from double scattering by cloud particles for materials with small $k \lesssim 0.1$, and from scattering by gaseous molecules first, followed by scattering by cloud particles for $k \gtrsim 0.1$. 
    
    Multiple scattering and the influence of gaseous molecules reduce the amplitude of $P_\mathrm{c}$. For planetary models that include cloud condensates with small $k$ such as $\mathrm{H_2O}$, $\mathrm{NH_3}$, or $\mathrm{KCl}$, characteristic features found in $P_\mathrm{c}$ for the second scattering order vanish in multiple scattering calculations. The resulting behavior is very predictable, with a single phase-angle-dependent change in the sign of $P_\mathrm{c}$ between $\alpha_\mathrm{p}$ of $90\degr$ and $100\degr$. Therefore, measuring $P_\mathrm{c}$ would allow for a characterization of the underlying asymmetry. In particular, it would allow one to determine which half of a planet is covered by a circumplanetary ring.
    
    Circular polarization phase curves of planetary models that include cloud particle species with high $k$, such as graphite, iron, $\mathrm{FeS}$, $\mathrm{FeO}$, $\mathrm{Fe_2O_3}$, or chromium, show distinct features caused by a change in the handedness of circular polarization at characteristic, phase-angle-dependent wavelengths when scattering by gaseous molecules before scattering by cloud particles is dominant. As the linear polarization behavior is very similar for all materials with $k \gtrsim 0.8$ \citep[see][]{lietzow2022}, precise measurements of circular polarization would allow for a better differentiation of these cloud particle species. As the wavelength dependence corresponds to a dependence on the size parameter $x_\mathrm{eff}$, it would also allow for a more accurate determination of the effective particle radius.

    In summary, the circular polarization of starlight reflected by giant exoplanets is sensitive to cloud particle composition and atmospheric asymmetries. It is typically three orders of magnitude smaller than the linear polarization, and therefore challenging to detect in unresolved observations due to stellar contamination \citep[see][]{wiktorowicz2025}. Nonetheless, it offers a promising complementary tool for atmospheric characterization with future high-precision instruments and coronagraphic observations of exoplanets.

\begin{acknowledgements}
We are grateful to the referee for a careful reading and insightful comments and suggestions that improved the paper.
This research made use of NASA’s Astrophysics Data System (\url{https://ui.adsabs.harvard.edu}), Astropy, a community-developed core Python package for Astronomy \citep{astropy2013,astropy2018,astropy2022}, Matplotlib (\url{https://matplotlib.org/}) \citep{hunter2007}, and NumPy (\url{https://numpy.org/}) \citep{harris2020}. It was supported through high-performance computing resources available at the Kiel University Computing Center. M.M. and S.W. thank the DFG for financial support under grant WO 857/24-1.
\newline\newline
Reproduced with permission from Astronomy \& Astrophysics, \copyright ESO
\end{acknowledgements}

\bibliography{references.bib}

\begin{appendix}

\nolinenumbers
\onecolumn
\section{Optical properties}
\label{app:optical_properties}
    \begin{figure*}[hb!]
        \centering
        \includegraphics[width = 0.94\linewidth]{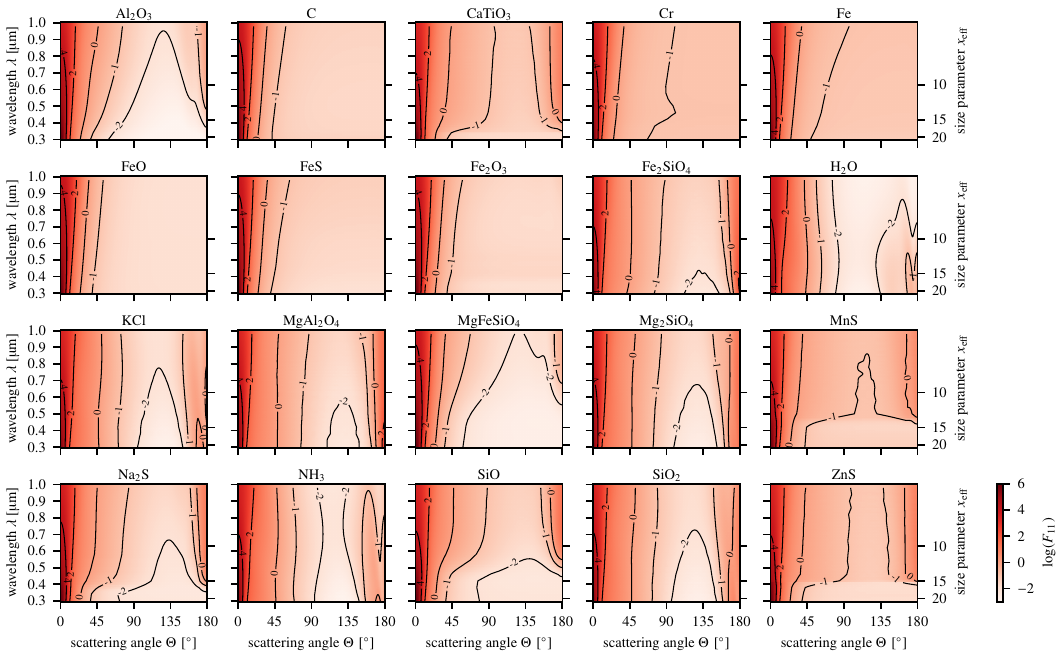}
        \caption{Logarithm of the scattering matrix element $F_{11}$ to base $10$ for various cloud condensates and wavelengths as a function of the single scattering angle $\Theta$ encoded in a color map with black contour lines.}
        \label{fig:F11}
        \includegraphics[width = 0.94\linewidth]{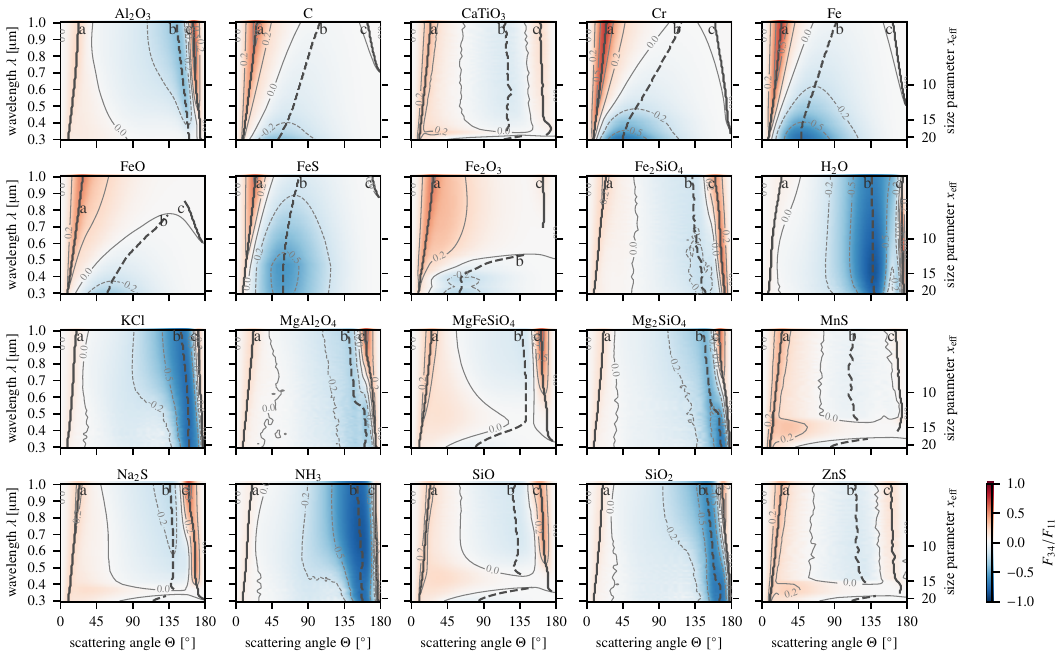}
        \caption{Ratio $F_{34}/F_{11}$ for various cloud condensates and wavelengths as a function of the single scattering angle $\Theta$ encoded in a color map. Light gray contour lines mark values of $F_{34}/F_{11}$. For each material, dark gray contour lines mark the location of the two highest positive peaks (a and c), and the highest negative peak (b), which are referenced in Sect. \ref{sec:semi_analytical_calculations}.}
        \label{fig:F34}
    \end{figure*}

    \twocolumn
    
   Here, we include visualizations of $\mathrm{log}\,F_{11}$ in Fig. \ref{fig:F11} 
   and of the ratio of the matrix elements $F_{34}$ and $F_{11}$ in Fig. \ref{fig:F34}. 
   Results of the single scattering linear polarization $P_\mathrm{s} = -F_{12}/F_{11}$ are presented in \citet{lietzow2022}. $P_\mathrm{s}$ is also visualized in Fig.~\ref{fig:mie_mie_scattering} with light gray contour lines.

\FloatBarrier
\section{Second scattering angle}
\label{app:second_scattering_angles}

    A sketch of the geometry is shown in Fig. \ref{fig:two_particle_model}.
    Here, $\hat{\vec{n}}_\mathrm{in}$ is the propagation direction of incoming radiation, $\hat{\vec{n}}_\mathrm{12}$ the propagation direction after the first scattering, and $\hat{\vec{n}}_\mathrm{out}$ the direction toward the observer.
    Incoming radiation moves in a positive direction along the $x$-axis. With the scattering angles $\theta_1, \phi_1$ of the first scattering, and the overall scattering angles $\theta, \phi$,
    \begin{equation}
        \hat{\vec{n}}_\mathrm{in} = 
        \begin{pmatrix}
            1 \\ 0 \\ 0
        \end{pmatrix}, \ 
        \hat{\vec{n}}_\mathrm{12} = 
        \begin{pmatrix}
            \cos\theta_1 \\
            \sin\theta_1 \cos\phi_1 \\
            \sin\theta_1 \sin\phi_1
        \end{pmatrix}, \ 
        \hat{\vec{n}}_\mathrm{out} = 
        \begin{pmatrix}
            \cos\theta \\
            \sin\theta \cos\phi \\
            \sin\theta \sin\phi \\
        \end{pmatrix}.
        \label{eq:vectors}
    \end{equation}
    From these definitions, the polar scattering angle $\theta_2$ and the azimuthal scattering angle $\phi_2$ of the second scattering are determined as functions of $\theta, \phi, \theta_1$, and $\phi_1$. As the radiation is moving in the direction of $\hat{\vec{n}}_\mathrm{12}$ before the second scattering and is scattered into the direction of $\hat{\vec{n}}_\mathrm{out}$ by the second particle,
    \begin{equation}
        \cos\theta_2 = \hat{\vec{n}}_\mathrm{12} \cdot \hat{\vec{n}}_\mathrm{out} = \sin\theta\sin\theta_1\cos(\phi_1 - \phi) + \cos\theta\cos\theta_1.
        \label{eq:cos_theta_2}
    \end{equation}
    $\theta_2$ is uniquely determined through this Equation.
    The azimuthal scattering angle $\phi_2$ is equal in size to the angle between the normals $\hat{\vec{n}}_\mathrm{1}$ and $\hat{\vec{n}}_\mathrm{2}$ to the two scattering planes measured around the rotation axis defined by $\hat{\vec{n}}_\mathrm{12}$ \citep[see, e.g.,][]{lietzow2021}. The scattering plane normals are
    \begin{equation}
        \hat{\vec{n}}_\mathrm{1} = \frac{\hat{\vec{n}}_\mathrm{12}\times\hat{\vec{n}}_\mathrm{in}}{\sin\theta_1},\ 
        \hat{\vec{n}}_\mathrm{2} = \frac{\hat{\vec{n}}_\mathrm{out}\times\hat{\vec{n}}_\mathrm{12}}{\sin\theta_2}.
        \label{eq:scattering_plane_normal}
    \end{equation}
    Both $\sin\phi_2$ and $\cos\phi_2$ are needed to uniquely determine $\phi_2$. 
    Following the approach by \citet{lietzow2021},
    \begin{align}
        \cos\phi_2 & = \hat{\vec{n}}_\mathrm{1} \cdot \hat{\vec{n}}_\mathrm{2}
        = \frac{\sin\theta\cos\theta_1\cos(\phi_1 -\phi) - \cos\theta\sin\theta_1}{\sin\theta_2},
        \label{eq:cos_phi_2} \\
        \sin\phi_2 & = (\hat{\vec{n}}_1 \times \hat{\vec{n}}_2) \cdot \hat{\vec{n}}_{12}
        = -\frac{\sin\theta\sin(\phi_1-\phi)}{\sin\theta_2}.
        \label{eq:sin_phi_2}
    \end{align}
    Note that it is possible to assume that the observer is located in the $xy$-plane at a positive $y$-coordinate, such that $\phi = 0$ (as in Sect. \ref{sec:semi_analytical_calculations}) without loss of generality. 

\FloatBarrier
\section{Antisymmetry of circular polarization}
\label{app:local_azimuthal_effect}

    \begin{figure}[!ht]
        \centering
        \includegraphics{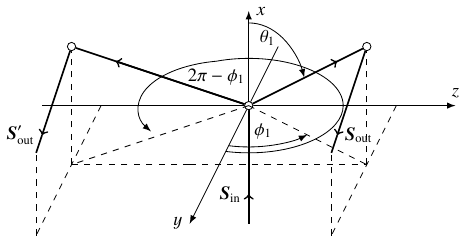}
        \caption{Symmetry of double scattering. The observer is located in the $xy$-plane at a positive $y$-coordinate. Possible positions of the two particles are shown. If the configuration is mirrored on the $xy$-plane by replacing $\phi_1$ with $2\pi-\phi_1$, the direction of circular polarization of the scattered radiation changes, but its absolute value remains the same.}
        \label{fig:symmetry}
    \end{figure}
    
    There is a local equivalent of the planetary polar effect.
    The component $V_\mathrm{out}$ of the Stokes vector $\vec{S}_\mathrm{out}$ in Eq. \eqref{eq:S_out_main} is
    \begin{equation}
        V_\mathrm{out}(\theta,\theta_1,\phi_1) = \left(\frac{\varpi}{4\pi}\right)^2 F_{34}(\theta_2)F_{12}(\theta_1)\sin(2\phi_2)\,I_0
        \label{eq:S_out}
    \end{equation}
    \citep[see also][]{kawata1978}.
    
    According to Eq. \eqref{eq:cos_theta_2}, \eqref{eq:cos_phi_2}, and \eqref{eq:sin_phi_2}, and assuming an overall azimuthal scattering angle of $\phi = 0$,
    \begin{align}
        \cos\theta_2(\theta,\theta_1,2\pi-\phi_1) & = \cos\theta_2(\theta,\theta_1,\phi_1), \notag \\
        \cos\phi_2(\theta,\theta_1,2\pi-\phi_1) & = \cos\phi_2(\theta,\theta_1,\phi_1), \notag \\
        \sin\phi_2(\theta,\theta_1,2\pi-\phi_1) & = -\sin\phi_2(\theta,\theta_1,\phi_1).
        \label{eq:trig_functions}
    \end{align}
    As $\sin(2\phi_2) = 2\sin\phi_2\cos\phi_2$, it follows that
    \begin{equation}
        \sin(2\phi_2(\theta,\theta_1,2\pi-\phi_1)) = -\sin(2\phi_2(\theta,\theta_1,\phi_1)).
        \label{eq:sin_2phi_2}
    \end{equation}
    Using Eq. \eqref{eq:S_out} and \eqref{eq:sin_2phi_2}, we derive Eq. \eqref{eq:anti-symmetry-V}
    \begin{equation*}
        V_\mathrm{out}(\theta,\theta_1,2\pi-\phi_1) = -V_\mathrm{out}(\theta,\theta_1,\phi_1).
    \end{equation*}
    Hence, the integral in Eq. \eqref{eq:anti-symmetry} is always zero. As this is ultimately caused by the dependence of $V_\mathrm{out}$ on the azimuthal scattering angles $\phi_1$ and $\phi_2$, this is the local azimuthal effect.
    
    A geometric argument is sketched in Fig. \ref{fig:symmetry}. If a considered configuration of the two particles is mirrored on the $xy$-plane, the circular polarization amplitude of the double-scattered radiation is constant while the direction of circular polarization changes.
    
\FloatBarrier
\section{Scattering in homogeneous semi-infinite atmospheres}
\label{app:infinitely_optically_thick}
    
    To calculate the Stokes vector of starlight reflected by a homogeneous semi-infinite planetary atmosphere on the circular polarization of the reflected starlight, we used a plane-parallel model of the planetary atmosphere. 
    Here, the aim is to derive an analytical description of the first and second scattering orders. The solution is used to interpret multiple-scattering results from high-performance numerical calculations with POLARIS in Sect. \ref{sec:results}.
 
    \begin{figure}[!tp]
        \centering
        \includegraphics{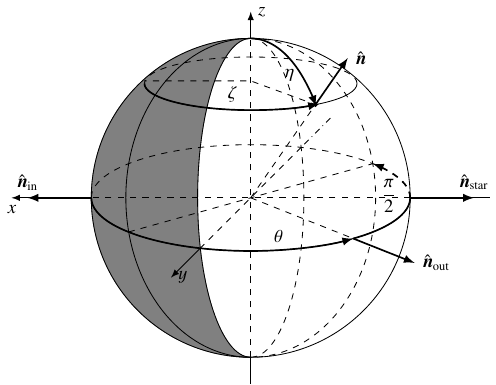}
        \caption{Definition of angles. The center of the planet is at the origin of the coordinate system. $\hat{\vec{n}}_\mathrm{star}$ points toward the star. $\hat{\vec{n}}_\mathrm{in}$ is the propagation vector of the radiation before the interaction with the planet. The great circle of the terminator is sketched. The night side of the planet is shown in gray. $\hat{\vec{n}}$ is the surface normal of a point on the upper edge of the planetary atmosphere with spherical coordinates $\eta$ and $\zeta$. The direction toward the observer is $\hat{\vec{n}}_\mathrm{out}$. $\theta$ is the overall scattering angle. A great circle marks the part of the planet that is visible to the observer.}
        \label{fig:planet_geometry}
    \end{figure}
    
    Consider a spherical planet with a single homogeneous atmospheric layer. The planetary center defines the center of the coordinate system. The center of the star is located on the negative $x$-axis at a large distance. Thus, 
    the direction from the planet toward the star is $\hat{\vec{n}}_\mathrm{star} = (-1,0,0)^T$.
    The direction from the planet toward the observer is $\hat{\vec{n}}_\mathrm{out} = (\cos\theta, \sin\theta, 0)^T$.
    The angles $\eta$ and $\zeta$ are spherical coordinates of the planet.
    The surface normal at a point on the top of the planetary atmosphere is
    \begin{equation}
        \hat{\vec{n}} =
        \begin{pmatrix}
            \sin\eta \cos\zeta \\
            \sin\eta \sin\zeta \\
            \cos\eta
        \end{pmatrix}.
        \label{eq:surface_normal}
    \end{equation}
    These definitions are visualized in Fig. \ref{fig:planet_geometry}.
    The local angles of incidence $\beta_\mathrm{in}$ and the local angle of reflectance $\beta_\mathrm{out}$ are \citep[equivalent equations are found in, e.g.,][]{horak1950,lester1979}
    \begin{align}
        \cos\beta_\mathrm{in} = \hat{\vec{n}} \cdot \hat{\vec{n}}_\mathrm{star} 
        & = -\sin\eta\cos\zeta \notag \\
        \cos\beta_\mathrm{out} = \hat{\vec{n}} \cdot \hat{\vec{n}}_\mathrm{out}
        & = \sin\eta\cos(\zeta-\theta).
        \label{eq:cos_beta_in_out}
    \end{align}
    If $\vec{S}(\theta,\eta,\zeta)$ is the Stokes vector of radiation that leaves the atmosphere at a point on the planetary atmosphere defined by $\eta$ and $\zeta$ in a direction defined by $\theta$, the total Stokes vector of one planetary hemisphere at an overall scattering angle of $\theta$ is
    \begin{equation}
        \vec{S}(\theta) = \frac{2}{\pi}\int_0^{\frac{\pi}{2}} \sin^2\eta
        \int_{\frac{\pi}{2}}^{\frac{\pi}{2}+\theta}
        \vec{S}(\theta,\eta,\zeta)\cos(\zeta-\theta)
        \,\mathrm{d}\zeta\,\mathrm{d}\eta.
        \label{eq:integration_equation}
    \end{equation}
    An equivalent equation for the flux density was derived by \citet{horak1950} for the entire planetary crescent.

\subsection{First scattering order}

\begin{figure}[!tp]
        \centering
        \includegraphics{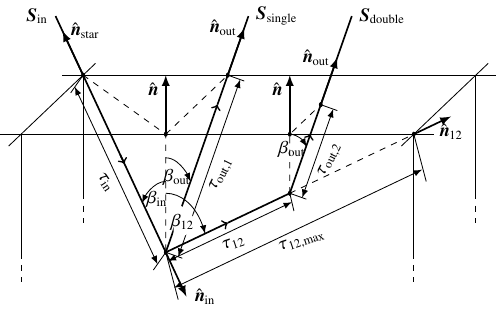}
        \caption{Light paths in a plane-parallel atmosphere. For a point on the upper edge of the atmosphere, $\hat{\vec{n}}$ is the surface normal, $\hat{\vec{n}}_\mathrm{star}$ is the direction toward the star, and $\hat{\vec{n}}_\mathrm{out}$ is the direction toward the observer. $\beta_\mathrm{in}$ is the angle of incidence. $\beta_\mathrm{out}$ is the angle of reflection. $\tau_\mathrm{in}, \tau_\mathrm{12}, \tau_\mathrm{out,1}, \tau_\mathrm{out,2}$, and $\tau_{12,\mathrm{max}}$ are the optical depths of the respective paths. $\vec{S}_\mathrm{in}$ is the Stokes vector of the incident radiation, and $\vec{S}_\mathrm{single}$ and $\vec{S}_\mathrm{double}$ are the Stokes vectors of single and double scattered radiation, respectively. For double-scattered radiation, $\hat{n}_{12}$ is the propagation direction after the second scattering event.}
        \label{fig:plane_parallel_geometry}
    \end{figure}

    The scattering geometry in the model atmosphere is sketched in Fig. \ref{fig:plane_parallel_geometry}. In a homogeneous atmosphere, the optical depth of a path is proportional to its length. Before the radiation is scattered for the first time, it reaches an optical depth of $\tau_\mathrm{in} > 0$. For the first scattering order, radiation is scattered directly toward the observer in the direction of $\hat{\vec{n}}_\mathrm{out}$. In that case, the radiation has to overcome an optical depth of
    \begin{equation}
        \tau_\mathrm{out,1} = \frac{\tau_\mathrm{in}\cos\beta_\mathrm{in}}{\cos\beta_\mathrm{out}} = -\frac{\tau_\mathrm{in}\cos\zeta}{\cos(\zeta-\theta)}
        \label{eq:tau_out_1}
    \end{equation}
    on the way out of the atmosphere.
    The contribution of such a path to the disk-integrated intensity is calculated using the element $F_{11}$ of the scattering matrix,
    the single scattering albedo $\varpi$ of the particles, the optical depth $\tau_\mathrm{in}$, and the expression for $\tau_\mathrm{out,1}$ in Eq. \eqref{eq:tau_out_1}. Overall,
    \begin{equation}
        I_\mathrm{s}  = -\frac{\varpi F_{11}(\theta)}{4\pi}\frac{I_0\cos\zeta}{\cos(\zeta-\theta)}\,\exp\left(-\tau_\mathrm{in}\left(1 - \frac{\cos\zeta}{\cos(\zeta-\theta)}\right)\right),
        \label{eq:I_single}
    \end{equation}
    where $I_0$ is the intensity of stellar radiation. For isotropic scattering, an equivalent expression is given by \citet{fairbairn2005}.
    To derive the total contribution of single scattering from a point on the upper edge of the planetary atmosphere, Eq. \eqref{eq:I_single} is integrated over the optical depth $\tau_\mathrm{in}$,
    \begin{equation}
        I_1(\theta,\eta,\zeta) = \int_0^\infty I_\mathrm{s}\,\mathrm{d}\tau_\mathrm{in}
        = \varpi \frac{F_{11}(\theta)}{4\pi}\frac{I_0\cos\zeta}{\cos\zeta - \cos(\zeta-\theta)}.
        \label{eq:I_one}
    \end{equation}
    For isotropic scattering, the Lommel-Seeliger law is extracted from this equation \citep[see][]{fairbairn2005,hapke1981}. In the limit of an infinite optical depth, results for single isotropic scattering by \citet{van_de_hulst1948} and for arbitrary scattering matrices and finitely optically thick plane-parallel atmospheres by \citet{hovenier1971} are equivalent to Eq. \eqref{eq:I_one}.
    
    Inserting $I_1(\theta,\eta,\zeta)$ into Eq. \eqref{eq:integration_equation} leads to an overall intensity of single scattered radiation of the upper half of the planet of
    \begin{align}
        I_1(\theta) = \varpi \frac{F_{11}(\theta)}{4\pi} I_0\cdot \frac{1}{2}
        \int_{\frac{\pi}{2}}^{\frac{\pi}{2}+\theta}\frac{\cos\zeta\cos(\zeta-\theta)}{\cos\zeta - \cos(\zeta-\theta)}
        \,\mathrm{d}\zeta.
        \label{eq:single_scattered_intensity}
    \end{align}
    The remaining integral is solved numerically in this study. 

\subsection{Second scattering order}

    In the case of the second scattering order, all possible first scattering directions have to be considered. After the first scattering event, the radiation now travels through the atmosphere along a different path in a direction described by the normalized vector $\hat{\vec{n}}_{12}$ as defined in Eq. \eqref{eq:vectors}. We define $\beta_{12}$ to be the angle between the surface normal $\hat{\vec{n}}$ and $\hat{\vec{n}}_\mathrm{12}$, so
    \begin{align}
        \cos\beta_{12} & = \sin\eta\cos\zeta\cos\theta_1 + \sin\eta\sin\zeta\sin\theta_1\cos\phi_1\notag \\
        &\hspace{0.5cm}+ \cos\eta\sin\theta_1\sin\phi_1.
        \label{eq:cos_beta_12}
    \end{align}
    The maximum optical depth $\tau_{12,\mathrm{max}}$ that radiation can reach in the direction of $\hat{\vec{n}}_{12}$ before the second scattering event is bounded by the upper border of the planetary atmosphere if $\beta_{12} \leq \pi/2$. Because $\cos\beta_{12} < 0$ if $\beta_{12} > \pi/2$, we have
    \begin{equation}
        \tau_{12,\mathrm{max}} = b\tau_\mathrm{in}, \quad b=
        \begin{cases}
            \cos\beta_\mathrm{in}/\cos\beta_{12}& \text{if } \cos\beta_{12} > 0, \\
            +\infty & \text{if } \cos\beta_{12} \leq 0.
        \end{cases}
        \label{eq:tau_12_max}
    \end{equation}
    The radiation is scattered toward the observer after it has traveled an optical path length of $\tau_\mathrm{12}$, where $0 \leq \tau_\mathrm{12} < \tau_\mathrm{12,max}$. Consequently, the radiation has to travel an optical depth of
    \begin{equation}
        \tau_\mathrm{out,2} = \frac{\tau_\mathrm{in}\cos\beta_\mathrm{in} - \tau_\mathrm{12}\cos\beta_{12}}{\cos\beta_\mathrm{out}}.
        \label{eq:tau_out}
    \end{equation}
    on its way out of the atmosphere. 
    The Stokes vector $\vec{S}_\mathrm{double}$ of double scattered radiation is calculated using Eq. \eqref{eq:S_out_main},
    \begin{equation}
        \vec{S}_\mathrm{double} = \vec{S}_\mathrm{out}\frac{\cos\beta_\mathrm{in}}{\cos\beta_\mathrm{out}}
         \exp(-\tau_\mathrm{in}-\tau_{12}-\tau_\mathrm{out,2}).
         \label{eq:S_double}
    \end{equation}
    Eq. \eqref{eq:S_double} is integrated on the optical depths $\tau_\mathrm{in}$ and $\tau_\mathrm{12}$ and on the scattering angles $\theta_1$ and $\phi_1$ to determine the total contribution $\vec{S}_2(\theta,\eta,\zeta)$ of double scattering from a point on the upper edge of the planetary atmosphere,
    \begin{align}
        \vec{S}_2(\theta,\eta,\zeta) =& 
         \int_0^{2\pi}\int_0^{\pi}\int_0^\infty\int_0^{\tau_{12,\mathrm{max}}} \vec{S}_\mathrm{double}
         \,\mathrm{d}\tau_\mathrm{12}\,\mathrm{d}\tau_\mathrm{in}
         \sin\theta_1\,\mathrm{d}\theta_1\,\mathrm{d}\phi_1 \notag \\
         & = \int_0^{2\pi}\int_0^\pi {\vec{S}_\mathrm{out}} f \sin\theta_1\,\mathrm{d}\theta_1\,\mathrm{d}\phi_1,
         \label{eq:S_2}
    \end{align}
    where $f$ is a function of $\theta,\theta_1,\phi_1,\eta$, and $\zeta$. However, it is more insightful to present $f$ in terms of the angles $\beta_\mathrm{in}, \beta_\mathrm{12}$, and $\beta_\mathrm{out}$. Using the parameter $b$ defined in Eq. \eqref{eq:tau_12_max},
    \begin{equation}
        f = \frac{\cos\beta_\mathrm{in}}{\cos\beta_\mathrm{out}-\cos\beta_{12}}\left(\frac{\cos\beta_\mathrm{out}}{\cos\beta_\mathrm{out} + \cos\beta_\mathrm{in}} - \frac{1}{1 + b(\beta_\mathrm{in},\beta_\mathrm{12})}\right).
        \label{eq:optical_depth_integration}
    \end{equation}
    From Eq. \eqref{eq:optical_depth_integration}, it follows that $f$ has a maximum for $\beta_{12} = \pi/2$. Thus, paths on which light travels parallel to the local upper border of the atmosphere between the two scattering events have the highest influence on the intensity $I_2(\theta,\eta,\zeta)$ for isotropic scattering. However, in our case, this effect is overpowered by the strongly forward-scattering nature of Mie scattering (Sect. \ref{sec:semi_analytical_calculations}).
    Again, the solutions by \citet{van_de_hulst1948} and \citet{hovenier1971} are equivalent to the equations above in the limit of an infinite optical depth. A more complete discussion of all possible cases for $\beta_\mathrm{in}$, $\beta_{12}$ and $\beta_\mathrm{out}$ was given by \citet{hovenier1971}.
    
    Inserting the Stokes vector $\vec{S}_2(\theta,\eta,\zeta)$ from Eq. \eqref{eq:S_2} into the integral in Eq. \eqref{eq:integration_equation} yields the Stokes vector of radiation that was scattered twice in the upper half of the planetary atmosphere before reaching the observer. The result of this calculation is of the form in Eq. \eqref{eq:double_scattered_stokes_main},
    \begin{equation*}
        \vec{S}_\mathrm{2}(\theta)
        = \int_0^{2\pi}\int_0^{\pi}
         {\vec{S}_\mathrm{out}}(\theta,\theta_1,\phi_1) \cdot P(\theta,\theta_1,\phi_1)
         \sin\theta_1\,\mathrm{d}\theta_1\,\mathrm{d}\phi_1,
    \end{equation*}
    where
    \begin{equation}
        P(\theta,\theta_1,\phi_1) = \frac{2}{\pi}\int_{0}^{\pi/2}\sin^2\eta\int_{\pi/2}^{\pi/2 + \theta} f\cos(\zeta-\theta)\,\mathrm{d}\zeta\,\mathrm{d}\eta
        \label{eq:P_definition}
    \end{equation}
    describes the influence of the planetary geometry.

\FloatBarrier
\section{Anti-symmetric part of P}
\label{app:probability_function}
    
    \begin{figure}
        \centering
        \includegraphics{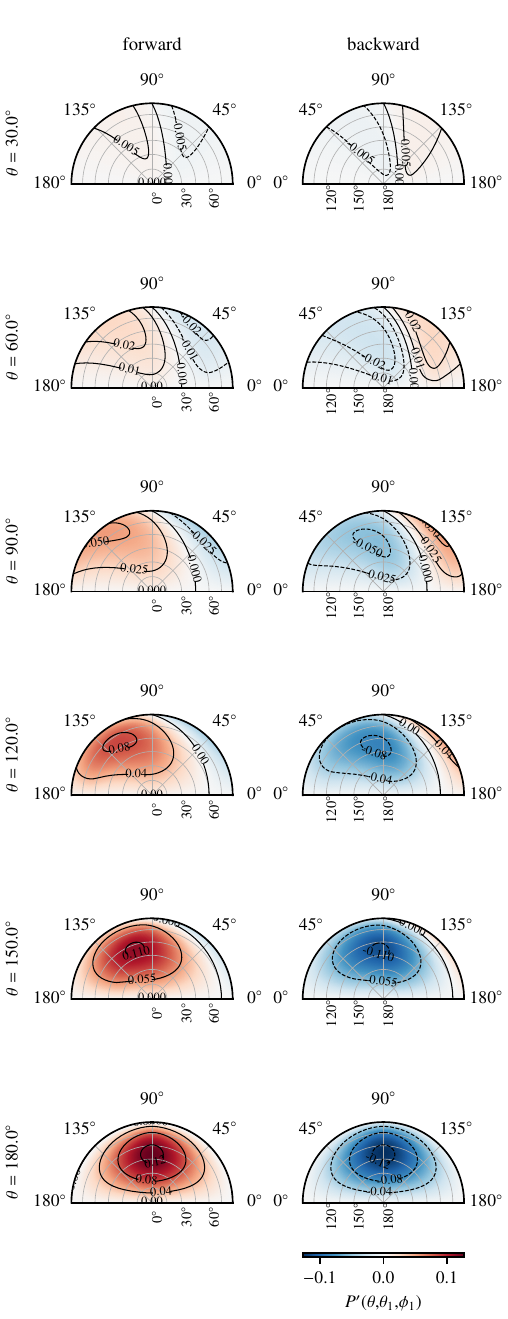}
        \caption{Color map of $P^\prime(\theta,\theta_1,\phi_1)$. Rows show different overall scattering angles $\theta$. Values of $\phi_1$ are given around the maps. Concentric gray half circles mark values of $\theta_1$ in $15\degr$ steps. Labels are positioned below and to the right of the intercept of their corresponding half circle with the $\phi = 0\degr$ line. Lambert azimuthal equal-area projections \citep[see, e.g.,][]{snyder1982} are used.}
        \label{fig:asymmetric_reduced}
    \end{figure}

    The resulting values of $P^\prime$ are visualized in Fig. \ref{fig:asymmetric_reduced}. Two types of contour lines where the value of $P^\prime$ is $0$ are identifiable. First, $P'(\theta,\theta_1,0)$ and $P'(\theta,\theta_1,\pi)$ are zero due to the definition of $P'$. Second, for any $\theta$, $P'(\theta,\theta_1,\phi_1)$ equals zero on the great circle defined as the intercept of the unit sphere with the plane $\tan(\theta/2) = \tan\theta_1\cos\phi_1$. This is due to an antisymmetry between sets of paths at different points on the upper edge of the planetary atmosphere. From Eq. \eqref{eq:cos_beta_in_out}, \eqref{eq:cos_beta_12}, and \eqref{eq:optical_depth_integration}, it follows that if $\tan(\theta/2) = \tan\theta_1\cos\phi_1$,
    \begin{align}
        &f(\theta, \theta_1,\phi_1,\eta,\zeta)\cos(\zeta-\theta) 
        \notag \\
        &= -f(\theta,\theta_1,2\pi - \phi_1,\eta,\pi - (\zeta-\theta))\cos\zeta.
        \label{eq:f_symmetry}
    \end{align}
    The geometry of this antisymmetry becomes more clear if $f$ is expressed as a function of $\beta_\mathrm{in}, \beta_\mathrm{out}$, and $\beta_\mathrm{12}$, 
    \begin{equation}
        f(\beta_\mathrm{in}, \beta_\mathrm{out}, \beta_\mathrm{12}) \cos\beta_\mathrm{out} = f(\beta_\mathrm{out}, \beta_\mathrm{in}, \pi -\beta_\mathrm{12})\cos\beta_\mathrm{in}.
        \label{eq:f_symmetry_insights}
    \end{equation}
    In a different mathematical framework, the same antisymmetry was described by \citet{hovenier1971}.

    \citeauthor{hovenier1971}’s antisymmetry divides the sphere of possible first scattering directions into two hemispheres with a positive and negative sign of $P^\prime$, respectively. The plane in which the previously discussed great circle lies further divides the overall scattering angle in half. 

\section{Different contributions}
\label{app:double_mie_scattering}

    \begin{figure*}[hb!]
        \centering
        \includegraphics{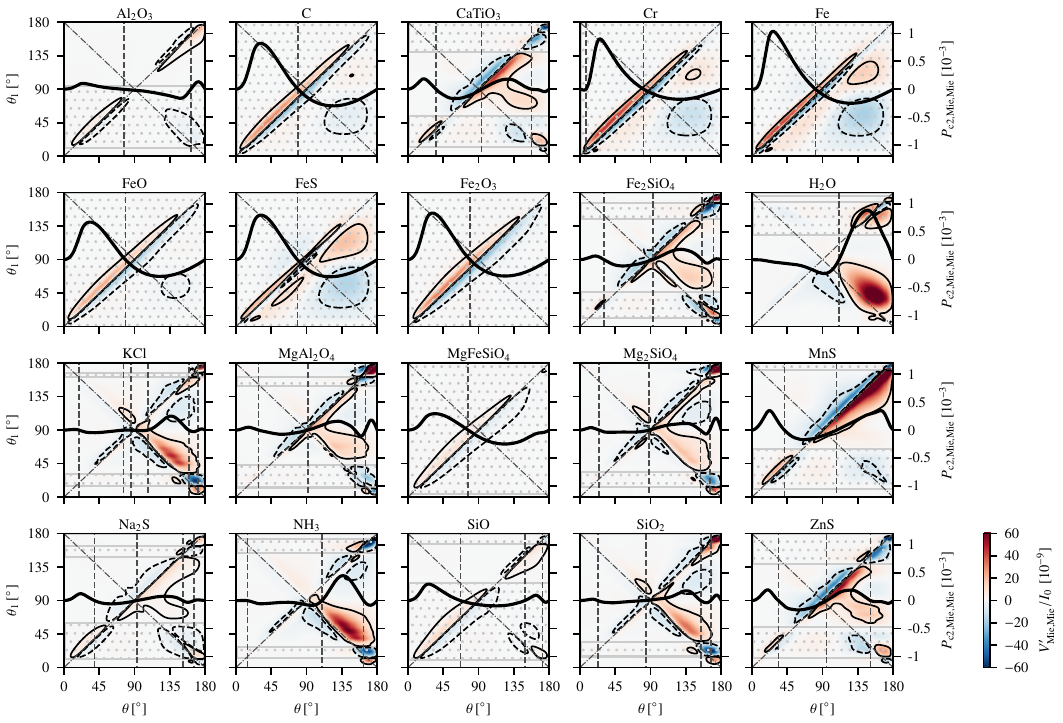}
        \caption{Color maps of the circularly polarized contribution $V^\prime_\mathrm{Mie,Mie}$ in units of incident stellar intensity $I_0$ of different first scattering angles $\theta_1$ to the circularly polarized intensity $V_\mathrm{2,Mie,Mie}$ of the second scattering order of a planetary hemisphere for various cloud condensates and a wavelength of $0.5\ \mathrm{\umu m}$. 
        A thick black line visualizes the overall circular polarization $P_\mathrm{c2,Mie,Mie}$ at $0.5\ \mathrm{\umu m}$. 
        Vertical dark gray lines mark zeros of $P_\mathrm{c2}$.
        Contour lines show the $85^\mathrm{th}$ percentile of the absolute value of $V^\prime_\mathrm{Mie,Mie}$ for each material.
        Areas marked with light gray dots mark $\theta_1$ where $-F_{12}/F_{11}$ is positive. 
        Dash-dotted diagonals mark $\theta = \theta_1$, which corresponds to scattering toward the observer first before scattering forward toward the observer \ref{item:forward_process}, and $\theta = 180\degr - \theta_1$, which corresponds to scattering away from the observer first and then backward toward the observer \ref{item:backward_process}.
        }
        \label{fig:mie_mie_features_500nm}
    \end{figure*}
    \begin{figure*}[ht!]
        \centering
        \includegraphics{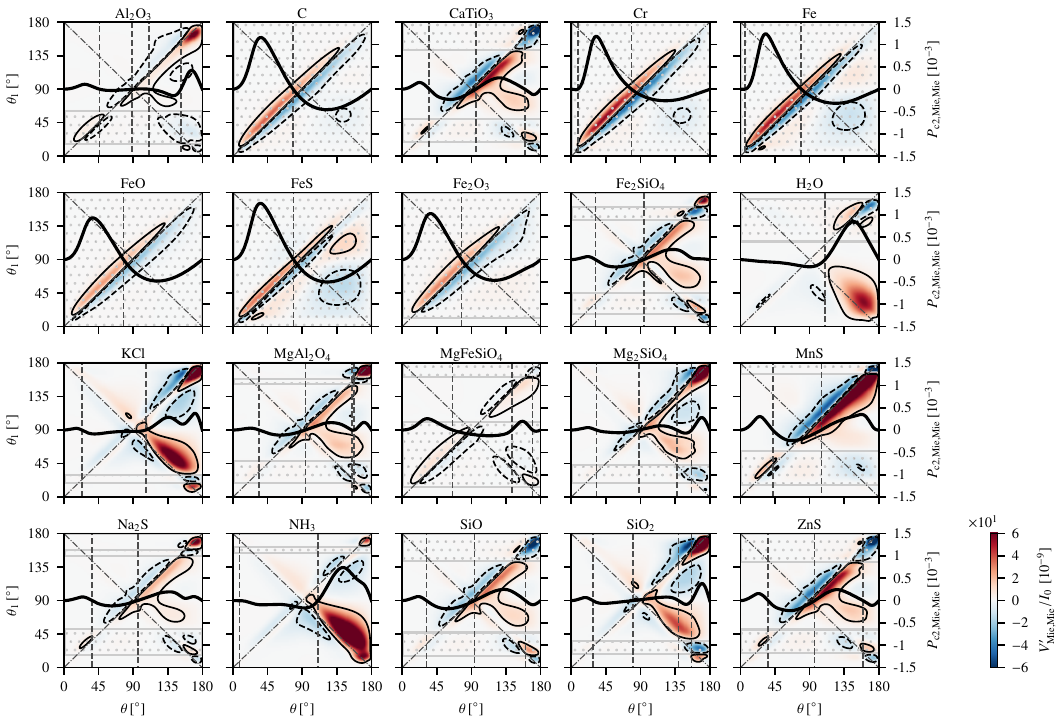}
        \caption{Similar to Fig. \ref{fig:mie_mie_features_500nm} but for a wavelength of $0.7\ \mathrm{\umu m}$.}
        \label{fig:mie_mie_features_700nm}
    \end{figure*}

    The contribution $V^\prime_\mathrm{Mie,Mie}(\theta,\theta_1)$ (Eq. \ref{eq:V_prime_definition}) to the circularly polarized flux of the second scattering order for double Mie scattering by cloud particles is visualized in Fig. \ref{fig:mie_mie_features_500nm} for a wavelength of $0.5\ \mathrm{\umu m}$ and in Fig. \ref{fig:mie_mie_features_700nm} for a wavelength of $0.7\ \mathrm{\umu m}$.

\subsection{Changes of handedness due to process (i)}
\label{app:process_i}
    
    In Fig. \ref{fig:mie_mie_features_500nm} and \ref{fig:mie_mie_features_700nm}, process \ref{item:forward_process} (introduced in Sect. \ref{sec:mie_mie_scattering}) corresponds to contributions close to the $\theta = \theta_1$ diagonal. For $\theta_1$ values that are larger than $\theta$, contributions to the circularly polarized flux $V^\prime$ with an identical sign to the single scattering linear polarization $P_\mathrm{s}(\theta_1)$ are found. If $\theta_1$ is smaller than $\theta$, $V^\prime$ has the opposite sign to $P_\mathrm{s}(\theta_1)$.
    
    In process \ref{item:forward_process}, $\theta_1$ differs from $\theta$ by a small characteristic angle $\Delta \theta_1$ of approximately $5\degr$ to $10\degr$ on average, while the average azimuthal scattering angle $\phi_1$ depends on $\theta$ but is typically small. If process \ref{item:forward_process} dominates, an approximation to the circularly polarized component $V_\mathrm{Mie,Mie}$ is therefore obtained by calculating the difference between the contribution of two sets of radiation paths characterized by ($\theta$, $\theta_1 = \theta + \Delta\theta_1$, $\phi_1$), and ($\theta$, $\theta_1 = \theta - \Delta\theta_1$, $\phi_1$), respectively. It is roughly proportional to
    \begin{align}
        V_{2,\mathrm{Mie,Mie}}(\theta) & \sim F_{12}(\theta+\Delta\theta_1)P^\prime(\theta,\theta+\Delta\theta_1,\phi_1)\sin(\theta+\Delta\theta_1) \notag \\
        & \hspace{0.5cm} - F_{12}(\theta-\Delta\theta_1)P^\prime(\theta,\theta-\Delta\theta_1,\phi_1)\sin(\theta-\Delta\theta_1).
    \end{align}
    This difference is developed into a Taylor series, resulting in
    \begin{equation}
        V_{2,\mathrm{Mie,Mie}}(\theta) \sim 2\Delta\theta_1\frac{\mathrm{d}}{\mathrm{d\theta}}(F_{12}(\theta)P^\prime(\theta,\theta,\phi_1)\sin\theta).
        \label{eq:derive_V_approx_from_F12}
    \end{equation}
    However, $P^\prime(\theta,\theta,\phi_1)$ is negative for all scattering angles if $\phi_1$ is small (see App. \ref{app:probability_function}) and the differential is dominated by the behavior of $F_{12}(\theta)\sin\theta$ for Mie scattering.
    Thus, $V_{2,\mathrm{Mie,Mie}}$ and similarly $P_\mathrm{c2,Mie,Mie}$ have zeros located in the vicinity of extrema of $F_{12}\sin\theta$, which are typically between two zeros of $P_\mathrm{s} = -F_{12}/F_{11}$ and close to extrema of $P_\mathrm{s}$.
    
    A sign change of circular polarization between two zeros of $F_{12}(\theta)\sin\theta$ further occurs only if these zeros are separated by more than a characteristic angle $2\Delta\theta_1$ of approximately $10\degr$ to $20\degr$. Otherwise, a sign change of $P_\mathrm{c2,Mie,Mie}$ expected from zeros of $F_{12}(\theta)\sin\theta$ is missing in $V_{2,\mathrm{Mie,Mie}}$.

\subsection{Relevance of the second scattering angle for iron}
\label{app:iron}

    \begin{figure}[ht!]
        \centering
        \vspace{1cm}
        \includegraphics{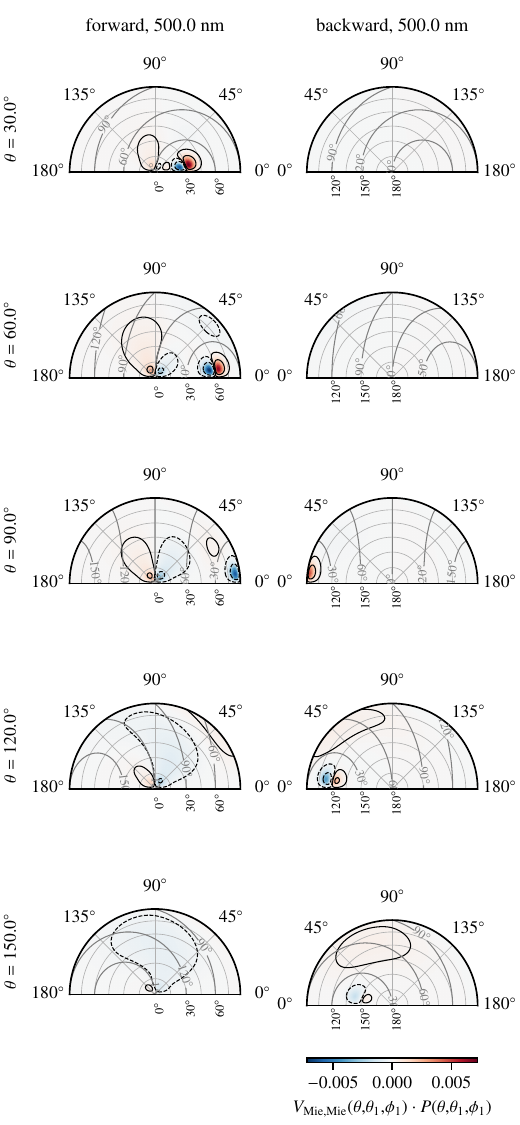}
        \caption{Color map of the quantity $V_\mathrm{Mie,Mie} P^\prime$ for double Mie scattering of radiation with a wavelength of $0.5\ \mathrm{\umu m}$ by iron particles. Angles are encoded similarly to Fig. \ref{fig:asymmetric_reduced}. Gray contour lines show values of $\theta_2$. Black contour lines mark the $80^\mathrm{th}$ and $99^\mathrm{th}$ percentile of the absolute value of $V_\mathrm{Mie,Mie} P^\prime$. Dashed lines correspond to negative values.}
        \vspace{1cm}
        \label{fig:iron}
    \end{figure}

    For the purpose of interpreting the influence of cloud particle species on the circular polarization of the second scattering order in Sect. \ref{sec:semi_analytical_calculations}, it is useful to visualize
    $
        V_{ij}(\theta,\theta_1,\phi_1) \cdot P^\prime(\theta,\theta_1,\phi_1),
    $
    which is the integrand in Eq. \eqref{eq:Vij_prime}. $V_{ij}P^\prime$ quantifies the contribution of the set of paths defined by $\theta$, $\theta_1$, and $\phi_1$ to the overall circularly polarized flux of the second scattering order $V_{2,ij}$.
    As an example, $V_\mathrm{Mie,Mie}P^\prime$ is shown for scattering by iron particles for a wavelength of $0.5\ \mathrm{\umu m}$ in Fig. \ref{fig:iron}. Gray contour lines show values of $\theta_2$ that are computed according to Eq. \eqref{eq:cos_theta_2}.

    The main contribution for iron is from process \ref{item:forward_process}, that is, scattering in a direction of $\theta_1 \approx \theta$ first. In Fig. \ref{fig:iron}, this corresponds to a positive and a negative feature of $V_\mathrm{Mie,Mie}P^\prime$ at small azimuthal scattering angles $\phi_1$. If $\theta_1$ is slightly smaller than $\theta$, negative polarization is induced. If $\theta_1$ is slightly larger than $\theta$, positive polarization is induced. Both features span a similar range $\Delta\theta_1$ of about $15\degr$. The second scattering is roughly forward by a scattering angle $\theta_2$ of less than $30\degr$. 
    
     At overall scattering angles $\theta$ above $90\degr$, two shallow features spread over a wide range of azimuthal scattering angles $\phi_1$. A positive feature is found for values of the overall scattering angle $\theta$ of $120\degr$ and $150\degr$ and for first scattering angles $\theta_1$ mostly larger than $90\degr$ in Fig. \ref{fig:iron}. These directions correspond to second scattering angles between $30\degr$ and $90\degr$. A negatively polarized feature is found at $\theta_1$ below $90\degr$ in directions corresponding to second scattering angles of more than $90\degr$. When integrated on $\phi_1$, these features result in the additional positive and negative features found in Fig. \ref{fig:mie_mie_features_500nm} (see also Sect. \ref{sec:high_k}).

     When $\theta$ is less than $90\degr$, an additional positive contribution appears for paths where $\theta_1$ is between $0\degr$ and $60\degr$, while $\phi_1$ is close to $90\degr$ (see Fig. \ref{fig:iron}). These contributions cause a faint positive feature in Fig. \ref{fig:mie_mie_features_500nm} for overall scattering angles $\theta$ below $90\degr$. This contribution is significantly more prominent for $\mathrm{FeS}$ at a wavelength of $0.5\ \mathrm{\umu m}$ in Fig. \ref{fig:mie_mie_features_500nm}.

\FloatBarrier
\subsection{Rayleigh-Mie scattering}
\label{app:H2_Mie_for_remaining_materials}

\begin{figure}[ht!]
        \centering
        \includegraphics{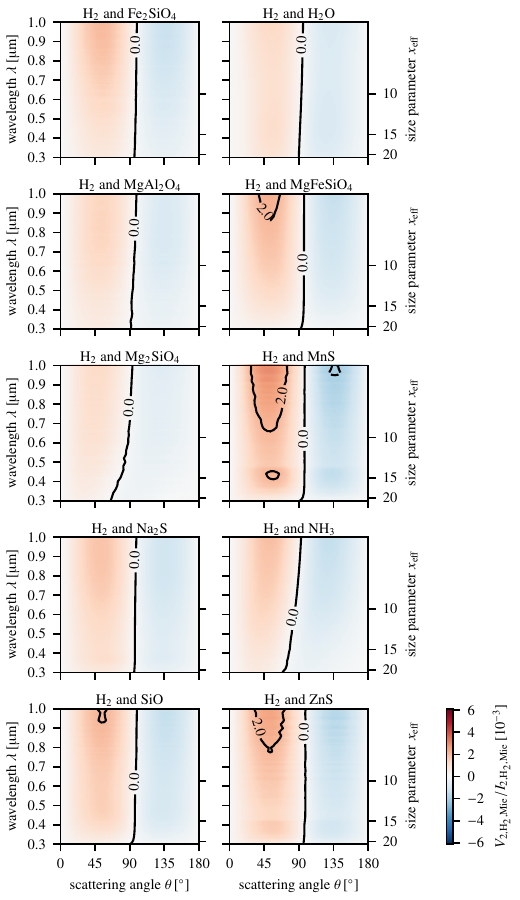}
        \caption{Similar to Fig. \ref{fig:rayleigh_mie_scattering}, but for different cloud condensates.}
        \label{fig:mixed_scattering_2}
    \end{figure}

\begin{figure*}
        \centering
        \includegraphics[width=\linewidth]{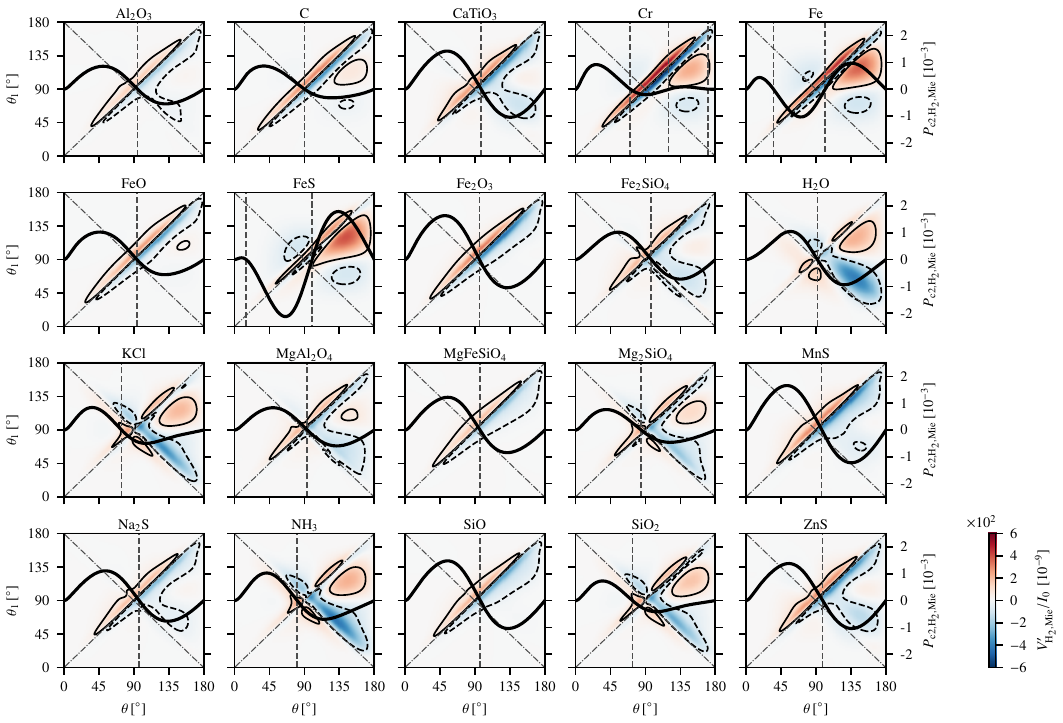}
        \caption{Color maps of the circularly polarized contribution $V^\prime_\mathrm{H_2,Mie}$ in units of incident stellar intensity $I_0$ of different first scattering angles $\theta_1$ to the circularly polarized intensity $V_\mathrm{2,H_2,Mie}$ of the second scattering order of a planetary hemisphere at a wavelength of $0.5\ \mathrm{\umu m}$. Contour lines show the $85^\mathrm{th}$ percentile of the absolute value of $V^\prime_\mathrm{H_2,Mie}$ for each material. A thick black line shows the overall circular polarization $P_\mathrm{c2,Mie,Mie}$ at $0.5\ \mathrm{\umu m}$.}
        \label{fig:H2_Mie_features}
    \end{figure*}

    The results for $P_\mathrm{c2,H_2,Mie}$ for the ten homogeneous mixtures of $\mathrm{H_2}$ molecules with cloud materials that are not included in Fig. \ref{fig:rayleigh_mie_scattering} are given in Fig. \ref{fig:mixed_scattering_2}. Features are very similar to $\mathrm{Al_2O_3}$ or $\mathrm{CaTiO_3}$ cloud particles in most cases. For forsterite particles, a similar behavior to $\mathrm{KCl}$ particles is found. 
    Results for $V^\prime_\mathrm{H_2,Mie}$ for all materials are found in Fig. \ref{fig:H2_Mie_features}.

\end{appendix}

\end{document}